\documentclass{sag00}

\begin{document}
\SetRunningHead{Y. Takeda}{Bahaviors of Ca II K line in A-type stars}
\Received{2020/10/31}
\Accepted{2020/11/23}

\title{Bahaviors of Ca II K line in A-type stars
}

%

\author{
Yoichi \textsc{Takeda}
}

\altaffiltext{}{11-2 Enomachi, Naka-ku, Hiroshima-shi, 730-0851, Japan }
\email{ytakeda@js2.so-net.ne.jp}
%


\maketitle

\begin{abstract}
A synthetic spectrum-fitting analysis was applied to the resonance Ca~{\sc ii} 
line at 3933.68~\AA\ for a large sample of 122 A-type main-sequence stars 
($7000 \ltsim T_{\rm eff} \ltsim 10000$~K) in a wide range of projected rotational 
velocity (10~$\ltsim v\sin i \ltsim 300$~km~s$^{-1}$), with an aim of investigating 
the behaviors of Ca abundances ([Ca/H]$_{39}$) determined from this strong 
Ca~{\sc ii} 3934 line, especially in context of (i) how they are related with the
Am phenomenon (often seen in slow rotators) and (ii) whether they are consistent 
with the Ca abundances ([Ca/H]$_{61}$) previously derived by the author from 
the weaker Ca~{\sc i} 6162 line. 
It was confirmed that Ca line strengths in Am stars tend to be weaker and 
associated abundances are lower compared to non-Am stars at the same $T_{\rm eff}$, 
indicating a deficiency of Ca in the photosphere of Am stars.
However, an appreciable fraction of cool Am stars ($T_{\rm eff} \ltsim 8000$~K)
were found to show extraordinarily anomalous Ca~{\sc ii} 3934 line feature (i.e.,  
unusually broad for its considerable weakness) which is hard to explain.
Regarding the comparison between [Ca/H]$_{39}$ and [Ca/H]$_{61}$, 
while both are roughly consistent with each other for hotter stars 
($T_{\rm eff} \gtsim 8000$~K), the former tends to be lower (by up to $\sim 1$~dex or 
even more) than the latter for cooler A stars ($T_{\rm eff} \ltsim 8000$~K) including 
those ``weak broad K line'' objects, This fact suggests that some special mechanism 
reducing the strength of Ca~{\sc ii} 3934 line is involved at $T_{\rm eff} \ltsim 8000$~K
where [Ca/H]$_{39}$ would be no more reliable. 
Whereas atomic diffusion causing the deficit of Ca in the photosphere as a result of
element segregation in the deeper radiative envelope may be regarded as a promising
explanation because it seems to fit in the qualitative trend of [Ca/H]$_{61}$ in A-type 
stars, the well-known feature of considerably weak Ca~{\sc ii} K line in classical Am 
stars should not necessarily be attributed to only this element diffusion scenario,    
for which some unknown weakening mechanism specific to this resonance line may 
independently be operative.
\end{abstract}

%


\section{Introduction}

The resonance line of once-ionized calcium at 3933.68~\AA (multiplet 1, historically
designated as ``K''), manifestly standing out in the violet region of the spectrum,
is so strong as to be visible in most stars from K through B type and is regarded
as one of the most important lines in stellar spectroscopy.
In context of early-type stars, this Ca~{\sc ii} K line is known to play a significant 
role in the spectral classification of chemically-peculiar stars, especially
with regard to metallic-lined A-type stars (hereinafter called Am stars), 
which occupy an appreciable fraction ($\sim$10--20\%) of A stars and are
generally slower rotators ($v\sin i \ltsim 100$~km~s$^{-1}$)  
characterized by strong metallic lines (e.g., those of Fe group elements).
That is, A-type stars are classified as Am, when three spectral types judged
from Ca~{\sc ii} K line, H~{\sc i} (Balmer) lines, and metallic lines are 
discrepant with each other as Sp(K)$<$Sp(H)$<$Sp(Metal); i.e., Ca~{\sc ii} K line
is weaker (indicating earlier type) while metallic lines are stronger (indicating
later type) compared to the spectral type from the Balmer lines (good metallicity-free 
indicator of $T_{\rm eff}$).

While such conspicuous weakness of Ca~{\sc ii} K line in Am stars was a matter of 
debate at first, it has become generally considered to be due to deficit of Ca in 
the atmosphere since the advent of diffusion theory in 1970s (e.g., Watson 1971; 
Smith 1973). That is, gravitational settling may take place in the deeper 
radiative envelope (possibly stable in slow rotators free from significant mixing), 
which would eventually yield surface underabundance of Ca (see, e.g., Richer et al. 
2000 or Talon et al. 2006 and the references therein for more recent theoretical
calculations).   

Actually, most chemical abundance studies on Am stars so far have shown 
the existence of an appreciable deficiency in Ca (along with Sc, another element 
showing weak lines in Am stars) by several tenths dex to $\sim 1$~dex 
(e.g., Conti 1970, Varenne \& Monier 1999, Hui-Bon-Hoa 2000, to mention a few). 
Quantitatively speaking, however, the extent of Ca abundance anomaly with regard  
to Am phenomena is rather diversified depending on each investigation; for example, 
Gebran et al. (2010) reported that Ca in Hyades Am stars is not so much deficient 
but even almost normal (cf. figure~5b therein).

Here, it should be pointed out that almost all these Ca abundance studies have 
been done by using lines of neutral calcium (a number of such Ca~{\sc i} lines 
are available in 4200--6500~\AA\ region; see, e.g., table~2 in Conti 1965). 
In contrast, systematic abundance investigations based on Ca~{\sc ii} 3934 
line for a large sample of upper main-sequence stars have rarely been 
done so far, which is presumably due to the considerable difficulty in precisely 
measuring the strength of this line with appreciable damping wings under the 
influence of blending with neighboring lines. Nevertheless, the following studies 
that investigated the Ca abundance trends for a number of A-type stars based 
on the Ca~{\sc ii} K line are worth mentioning.
\begin{itemize}
\item
Henry (1969) carried out narrow-band photometric observations of Ca~{\sc ii} 3934 
line ($k$ index) for an extensive sample of 146 A-type stars and investigated 
the strengths of this line. He found that the Ca abundances of sharp-line stars 
vary by a factor of $\sim 2$ and (somewhat surprisingly) that a substantial 
fraction of Am stars have normal K-line strengths. 
\item
Guthrie (1987) derived the equivalent widths of Ca~{\sc ii} 3934 line for 57 
Am stars mainly based on the published $k$ indices or K-line spectral types 
taken from various literature, 
and determined their Ca abundances, which showed a large dispersion but were 
generally underabundant relative to the Sun ($-1 \ltsim$~[Ca/H]~$\ltsim 0$).  
The deficiency appeared to have some correlation with age (i.e., position on the 
HR diagram), which he suggested may be consistent with the diffusion theory. 
\item
North et al. (1997), in their study of absolute magnitude of chemically peculiar
stars, mentioned their Ca abundance determinations for 27 Am stars from the 
Ca~{\sc ii} K line by using the spectrum synthesis method based on their 
high-dispersion spectroscopic observations. Since their abundances were
in good agreement with Guthrie's (1987) results for 8 stars in common, 
they combined both to make a large data sample of 76 Am stars, from which 
they found a $T_{\rm eff}$-dependence of Ca abundances (i.e., deficiency
tends to be more enhanced towards lower $T_{\rm eff}$). Unfortunately, 
they neither gave any account about the procedure of their spectrum 
synthesis analysis nor detailed abundance results for scrutiny. 
\end{itemize}

As such, comprehensive Ca abundance studies of A and Am stars based on the 
Ca~{\sc ii} K line of ionized calcium are few in number, despite that
several investigations using Ca~{\sc i} lines of neutral calcium using 
high-quality are already available. Given that suspicion exited in the past 
(e.g., B\"{o}hm-Vitense 1976, 2006) that the weakness of Ca~{\sc ii} K line 
in Am stars might be due to unusual atmospheric condition (e.g., anomalous 
ionization) rather than the abundance effect, careful analysis of 
the Ca~{\sc ii} 3934 line on a large sample of A-type stars would be in order 
and worthwhile.

About a decade ago, Takeda et al. (2009; hereinafter referred to as Paper~I) 
conducted Na abundance determinations based on the synthetic spectrum fitting 
method applied to the 5880--5905~\AA\ region comprising Na~{\sc i} 5890/5896 
doublet (D) for 122 A-type stars of slow as well as rapid rotators 
(10~$\ltsim v\sin i \ltsim$~300~km~s$^{-1}$). 
Ca abundances (along with those of O, Fe, Si, and Ba) were also 
determined for these sample stars, since a supplementary spectrum fitting 
analysis was likewise carried out in that study on the 6140--6170~\AA\ region 
(for the purpose of determining $v\sin i$ and abundances of important key elements)
which comprises Ca~{\sc i} lines (among which the strongest is the Ca~{\sc i} 
6162.17 line of multiplet 3).
Fortunately, thanks to the wide wavelength coverage ($\sim$~3700--10000~\AA) 
of the echelle spectra employed in Paper~I, they can be used also for the 
analysis of Ca~{\sc ii} 3934 line of present interest. 

Accordingly, the author thought about conducting Ca abundance determinations from 
the Ca~{\sc ii} K line for these 122 stars based on the similar spectrum fitting 
analysis applied to the observational data used in Paper~I,
while placing the main focus on the following questions: (i) How are the Ca 
abundances established from the Ca~{\sc ii} K line related with Am 
phenomenon? Is there a distinct difference between normal A stars and Am 
stars in this respect? (ii) Are the Ca abundances derived from the Ca~{\sc i} 6162 line
in Paper~I and those from the Ca~{\sc ii} 3934 line consistent with each other?
This comparison may serve for checking the usability of Ca~{\sc ii} K line as 
an abundance indicator. To clarify these points was the purpose of this investigation. 

\section{Observational data}

The program stars of this study are 122 apparently bright ($V \ltsim 6$) 
A-type main-sequence stars, which are listed in table~1.   
The basic observational data of these targets are the echelle spectra (with a 
spectral resolution of $R \sim 45000$) widely covering the wavelength range 
from $\sim 3700$~\AA\ to $\sim 10000$~\AA, which were obtained 
in 2008 January, 2008 September, and 2009 January by using BOES 
(Bohyunsan Observatory Echelle Spectrograph) 
attached to the 1.8 m reflector at Bohyunsan Optical Astronomy Observatory. 
See section~2 in Paper~I for more details regarding the observations and data 
reduction. Due to the lowered sensitivity of the CCD detector in the violet 
region, the S/N ratio around the Ca~{\sc ii}~3934 line of present interest 
is not so high (from several tens to $\sim$~100--200 differing from star 
to star) as compared to the yellow--red region.
The violet-region spectrum of Procyon, which was employed as the standard 
reference star, was adopted from the UVES-POP database (Bagnulo et al. 2003).\footnote{
http://www.eso.org/sci/observing/tools/uvespop/bright\_stars\_uptonow.html}

\section{Stellar parameters}

Regarding the fundamental atmospheric parameters of 122 program stars, 
the same values as used in Paper~I were adopted (see subsection~3.1
therein for more details): The effective temperature ($T_{\rm eff}$) and 
the surface gravity ($\log g$) were determined photometrically mainly 
from Str\"{o}mgren's $uvby\beta$ colors according to Napiwotzki et al.'s (1993) 
calibration, while the microturbulence ($v_{\rm t}$) was derived from 
$T_{\rm eff}$ by using Takeda et al.'s (2008) empirical formula.  
These parameters are summarized in table~1.
Among the 122 sample stars, 28 are metallic-lined A-type (Am) stars, which 
are classified as Am in (at least one of) the published spectral classifications 
compiled in the SIMBAD database (cf. column 16 in table~1).
As in Paper~I, Kurucz's (1993) solar-metallicity model-atmosphere grid was 
used to be interpolated in terms of $T_{\rm eff}$ and $\log g$ for each star.
Regarding Procyon, the same atmospheric parameters and model atmosphere 
as adopted in Paper~I were used (cf. section IV(c) in Takeda et al. 2008).  

\setcounter{table}{0}
\scriptsize
\renewcommand{\arraystretch}{0.8}
\setlength{\tabcolsep}{3pt}
\begin{longtable}{crrcc ccccc cr cr c l}
\caption{Basic stellar data and the results of the analysis.}
\hline\hline
star\# & HD\# & HR\# & Name & Sp.type & $T_{\rm eff}$ & $\log g$ & $v_{\rm t}$ & 
$v\sin i_{61}$ & [Fe/H]$_{61}$ & [Ca/H]$_{61}$ & $W_{61}$ & [Ca/H]$_{39}$ & $W_{39}$ & 
$\alpha$ & Remark \\
\hline
\endhead
\hline
\endfoot
\hline
\multicolumn{16}{l}{\hbox to 0pt{\parbox{150mm}{\footnotesize
In columns 1--8 are given the star number (arbitrarily assigned in this study), 
HD number, HR number, star name (with constellation), spectral type (taken from 
the Bright Star Catalogue; Hoffleit \& Jaschek 1982), effective temperature (in K), 
logarithmic surface gravity (in  cm~s$^{-2}$/dex), and microturbulent velocity 
(in km~s$^{-1}$), which are the same as in table~1 of Paper~I.
Columns 9 through 12 present the results (designated by suffix ``61'')
obtained from the 6140--6170 \AA\ region fitting carried out in Paper~I: 
the projected rotational velocity (in km~s$^{-1}$), Fe abundance, 
Ca abundance and the equivalent width of Ca~{\sc i} 6162 line. 
The Ca abundance and equivalent width of the Ca~{\sc ii} 3934
line (designated by suffix ``39'') obtained from the 3910--3955 \AA\ region 
fitting analysis of this study are given in columns 13 and 14, respectively.
Column 15 presents $v\sin i_{39}/v\sin i_{61}$ ($\equiv \alpha$),
which is the ratio of $v\sin i$ values derived from two fitting analyses of 
3910--3955 \AA\ region and 6140--6170 \AA\ region.
In the final column 16, ``Am'' means stars classified as metallic-line star in 
(at least one of) the published spectral classifications compiled in the 
SIMBAD database (note that this does not necessarily consistent with the spectral 
type given in column 5), and ``WBK'' denotes Weak Broad K-line star (defined by
the criterion $\alpha > 1.5$). 
All the abundance results ([X/H]; in dex) are the differential values relative to 
Procyon ($A_{61}^{\rm Fe} = 7.49$, $A_{61}^{\rm Ca} = 6.19$, and $A_{39}^{\rm Ca} = 6.15$,
which are almost equivalent to the solar abundances) and equivalent widths are in m\AA. 
The parenthesized values in columns 13 and 14 are unreliable and thus should not be
seriously taken because they are of the anomalous Weak Broad K-line stars.
The 122 stars are arranged in the descending order of $v_{\rm e} \sin i_{61}$,
to make it consistent with figure~1 (and figure~2 in Paper~I). 
\\ \\
}}}
\endlastfoot
\hline
   1& 130109&5511&109 Vir&A0V            &   9683&  3.68&  2.4&   290& $-$0.80 & $\cdots$ &  $\cdots$& +0.24 &  985 & 1.21 &       \\
   2& 192696&7740&33 Cyg&A3IV-Vn         &   7815&  3.49&  4.0&   268& $-$0.30 & +0.19 & 164 & $-$0.79 & 1741 & 0.87 &       \\
   3&  38678&1998&$\zeta$ Lep&A2Vann     &   8610&  3.96&  3.7&   246& $-$0.48 & $-$0.09 &  69 & $-$0.41 & 1404 & 0.99 &       \\
   4&  28024&1392&$\upsilon$ Tau&A8Vn    &   7107&  3.20&  3.3&   233& $-$0.05 & +0.01 & 167 & $-$0.56 & 3793 & 0.72 &       \\
   5&  19275& 932&         &A2Vnn        &   9111&  4.12&  3.2&   232& $-$0.16 & $-$0.42 &  18 & $-$0.06 & 1358 & 1.26 &       \\
   6& 187642&7557&$\alpha$ Aql&A7IV-V    &   7717&  4.00&  3.9&   231& +0.19 & $-$0.51 & 104 & $-$0.37 & 2899 & 0.85 &       \\
   7& 203280&8162&$\alpha$ Cep&A7IV-V    &   7585&  3.73&  3.8&   229& +0.14 & +0.28 & 185 & $-$0.36 & 3280 & 0.86 &       \\
   8&  92769&4189&40 LMi&A4Vn            &   7820&  4.01&  4.0&   225& $-$0.06 & +0.05 & 156 & $-$0.79 & 1666 & 0.92 &       \\
   9&  14055& 664&$\gamma$ Tri&A1Vnn     &   9335&  3.98&  2.9&   221& +0.00 & $\cdots$ &  $\cdots$& $-$0.33 &  819 & 1.13 &       \\
  10& 118098&5107&$\zeta$ Vir&A3V        &   8249&  4.02&  4.0&   218& +0.16 & $\cdots$ &  $\cdots$& $-$0.58 & 1565 & 0.93 &       \\
  11&  33111&1666&$\beta$ Eri&A3IIIvar   &   7928&  3.59&  4.0&   217& +0.02 & +0.39 & 175 & $-$0.62 & 1910 & 0.83 &       \\
  12& 116842&5062&80 UMa&A5V SB          &   7942&  4.02&  4.0&   217& $-$0.10 & $-$0.09 & 134 & $-$0.32 & 2600 & 0.96 &       \\
  13& 106591&4660&$\delta$ UMa&A3Vvar    &   8629&  3.85&  3.7&   210& $-$0.38 & $-$0.31 &  46 & $-$0.15 & 1817 & 1.05 &       \\
  14&  19107& 925&$\rho^{3}$ Eri&A8V     &   7772&  3.96&  4.0&   197& +0.00 & +0.01 & 155 & $-$0.96 & 1438 & 0.99 &       \\
  15&  37507&1937&49 Ori&A4V             &   7979&  3.82&  4.0&   192& $-$0.06 & +0.03 & 140 & $-$0.81 & 1484 & 1.01 &       \\
  16& 126248&5392&         &A5V          &   8212&  4.02&  4.0&   190& $-$0.15 & $-$0.35 &  83 & $-$1.05 &  957 & 1.09 &       \\
  17&  79439&3662&18 UMa&A5V             &   7822&  4.03&  4.0&   187& +0.08 & +0.07 & 158 & $-$0.44 & 2478 & 0.90 &       \\
  18&  27946&1388&$\kappa^{2}$ Tau&A7V   &   7401&  3.84&  3.7&   187& $-$0.06 & $-$0.32 & 136 & $-$0.29 & 3865 & 0.90 &       \\
  19& 197950&7945&4 Cep&A8V              &   7768&  4.08&  4.0&   186& +0.20 & +0.02 & 157 & $-$0.27 & 3139 & 0.89 &       \\
  20&  97603&4357&$\delta$ Leo&A4V       &   8180&  3.90&  4.0&   184& +0.06 & +0.26 & 148 & $-$0.16 & 2636 & 0.91 &       \\
  21& 141003&5867&$\beta$ Ser&A3V        &   8580&  3.56&  3.7&   180& $-$0.47 & $-$0.85 &  15 & $-$0.25 & 1662 & 1.06 &       \\
  22&  16970& 804&$\gamma$ Cet&A3V       &   9122&  4.05&  3.2&   180& +0.08 & +0.14 &  45 & +0.16 & 1704 & 0.96 &       \\
  23& 102124&4515&$\xi$ Vir&A4V          &   8026&  4.09&  4.0&   177& +0.24 & +0.14 & 152 & $-$0.34 & 2395 & 0.80 &       \\
  24&  18978& 919&$\tau^{3}$ Eri&A4V     &   8062&  4.03&  4.0&   175& +0.09 & +0.22 & 155 & $-$1.05 & 1067 & 0.98 &       \\
  25&  59037&2857&64 Gem&A4V             &   8238&  3.99&  4.0&   173& $-$0.07 & $-$0.13 & 105 & $-$0.18 & 2484 & 0.95 &       \\
  26& 192425&7724&$\rho$ Aql&A2V         &   8984&  4.21&  3.3&   167& +0.18 & $-$0.19 &  37 & $-$0.19 & 1337 & 0.93 &       \\
  27&  30780&1547&97 Tau&A7IV-V          &   7644&  3.87&  3.9&   165& $-$0.07 & $-$0.12 & 147 & $-$0.70 & 2100 & 1.37 &       \\
  28&  80081&3690&38 Lyn&A1V             &   9014&  3.82&  3.3&   163& $-$0.36 & +0.04 &  39 & +0.09 & 1657 & 1.21 &       \\
  29& 110411&4828&$\rho$ Vir&A0V         &   9117&  4.22&  3.2&   162& $-$0.47 & $-$0.61 &  13 & $-$1.26 &  412 & 1.02 &       \\
  30& 103287&4554&$\gamma$ UMa&A0V SB    &   9202&  3.79&  3.0&   161& $-$0.21 & $-$0.42 &  11 & $-$0.26 &  950 & 1.05 &       \\
  31&   6695& 328&$\psi^{2}$ Psc&A3V     &   8765&  4.13&  3.6&   156& $-$0.12 & +0.23 &  89 & +0.07 & 2147 & 0.92 &       \\
  32&  56537&2763&$\lambda$ Gem&A3V...   &   8458&  3.90&  3.8&   148& $-$0.10 & +0.14 & 106 & +0.00 & 2516 & 0.96 &       \\
  33& 135559&5679&4 Ser&A4V              &   7992&  4.14&  4.0&   145& $-$0.14 & $-$0.21 & 119 & $-$1.26 &  889 & 1.31 &       \\
  34&  20677&1002&32 Per&A3V             &   8952&  4.08&  3.3&   142& +0.04 & $-$0.05 &  46 & $-$0.10 & 1490 & 0.92 &       \\
  35& 210418&8450&$\theta$ Peg&A2V       &   8888&  3.82&  3.4&   141& $-$0.43 & $-$0.20 &  33 & $-$0.17 & 1388 & 1.00 &       \\
  36&  87696&3974&21 LMi&A7V             &   7878&  4.13&  4.0&   140& $-$0.11 & $-$0.15 & 133 & $-$0.37 & 2569 & 1.09 &       \\
  37&   8538& 403&$\delta$ Cas&A5Vv SB   &   7776&  3.41&  4.0&   138& $-$0.16 & $-$0.19 & 128 & $-$0.52 & 2468 & 0.77 &       \\
  38&  76644&3569&$\iota$ UMa&A7IV       &   7934&  4.22&  4.0&   135& $-$0.13 & +0.05 & 151 & $-$0.07 & 3485 & 1.01 &       \\
  39&  91312&4132&         &A7IV         &   7724&  4.08&  3.9&   133& $-$0.10 & $-$0.03 & 152 & $-$0.59 & 2234 & 1.38 &       \\
  40&  29488&1479&$\sigma^{2}$ Tau&A5Vn  &   7990&  3.82&  4.0&   132& $-$0.08 & $-$0.04 & 132 & $-$0.30 & 2610 & 0.82 &       \\
  41& 222661&8988&$\omega^{2}$ Aqr&B9V   &  10481&  4.28&  1.5&   130& $-$0.13 & $-$0.40 &   1 & +0.05 &  509 & 1.11 &       \\
  42&  71155&3314&         &A0V          &   9718&  4.11&  2.4&   127& $-$0.39 & $-$0.61 &   3 & $-$0.36 &  584 & 1.12 &       \\
  43&  99211&4405&$\gamma$ Crt&A9V       &   7722&  3.95&  3.9&   124& $-$0.16 & $-$0.19 & 137 & $-$0.23 & 3383 & 1.00 &       \\
  44& 218045&8781&$\alpha$ Peg&B9.5III   &   9643&  3.52&  2.5&   121& $-$0.34 & $\cdots$ &  $\cdots$& $-$0.16 &  637 & 1.14 &       \\
  45&   1404&  68&$\sigma$ And&A2V       &   8828&  4.00&  3.5&   121& $-$0.29 & $-$0.23 &  40 & $-$0.18 & 1501 & 0.96 &       \\
  46& 139006&5793&$\alpha$ CrB&A0V       &   9573&  3.87&  2.5&   121& $-$0.16 & +0.11 &  15 & $-$0.06 &  852 & 1.04 &       \\
  47& 213558&8585&$\alpha$ Lac&A1V       &   9434&  4.14&  2.7&   119& $-$0.21 & $-$0.38 &  10 & +0.01 & 1112 & 1.12 &       \\
  48& 127762&5435&$\gamma$ Boo&A7IIIvar  &   7663&  3.59&  3.9&   117& $-$0.23 & $-$0.12 & 143 & $-$0.26 & 3518 & 1.02 &       \\
  49& 102647&4534&$\beta$ Leo&A3Vvar     &   8643&  4.17&  3.7&   117& $-$0.07 & +0.04 &  85 & $-$0.08 & 2018 & 0.97 &       \\
  50&  28527&1427&         &A6IV         &   8039&  3.99&  4.0&   116& +0.01 & $-$0.01 & 134 & $-$0.05 & 3338 & 0.88 &       \\
  51&  32301&1620&$\iota$ Tau&A7V        &   7937&  3.74&  4.0&   116& $-$0.07 & $-$0.02 & 136 & $-$0.14 & 3271 & 0.97 &       \\
  52& 177196&7215&16 Lyr&A7V             &   7940&  4.10&  4.0&   116& $-$0.09 & $-$0.16 & 128 & $-$0.22 & 2929 & 1.01 &       \\
  53&  28910&1444&$\rho$ Tau&A8V         &   7520&  3.97&  3.8&   111& $-$0.24 & $-$0.43 & 124 & $-$0.30 & 3621 & 1.25 &       \\
  54& 137898&5746&10 Ser&A8IV            &   7582&  3.97&  3.8&   109& $-$0.05 & $-$0.10 & 152 & ($-$0.99) & (1569) & 1.77 &WBK  \\
  55&  48097&2466&26 Gem&A2V             &   8984&  4.23&  3.3&   109& $-$0.12 & $-$0.04 &  48 & $-$0.24 & 1275 & 1.00 &       \\
  56& 200761&8075&$\theta$ Cap&A1V       &   9633&  4.11&  2.5&   106& +0.02 & +0.26 &  23 & +0.06 &  988 & 1.02 &       \\
  57&  31295&1570&$\pi^{1}$ Ori&A0V      &   8993&  4.11&  3.3&   105& $-$0.72 & $-$0.91 &   8 & $-$0.96 &  576 & 1.10 &       \\
  58& 125162&5351&$\lambda$ Boo&A0sh     &   8834&  4.08&  3.5&   103& $-$1.54 & $\cdots$ &  $\cdots$& $-$1.74 &  343 & 1.06 &       \\
  59&  85376&3900&22 Leo&A5IV            &   7459&  3.98&  3.7&   102& $-$0.18 & $-$0.50 & 117 & $-$0.91 & 1851 & 1.31 &       \\
  60&  31647&1592&$\omega$ Aur&A1V       &   9478&  4.27&  2.7&   101& +0.00 & $-$0.13 &  17 & $-$0.22 &  871 & 1.08 &       \\
  61&   6961& 343&$\theta$ Cas&A7Vvar    &   7900&  3.81&  4.0&   100& +0.07 & $-$0.09 & 133 & $-$0.26 & 2908 & 0.96 &       \\
  62&  18454& 883&4 Eri&A5IV/V           &   7740&  4.07&  3.9&   100& +0.21 & +0.05 & 159 & $-$0.26 & 3234 & 1.17 &       \\
  63&  76543&3561&$o^{1}$ Cnc&A5III      &   8330&  4.18&  3.9&    91& +0.31 & +0.05 & 117 & $-$0.27 & 2075 & 1.25 &       \\
  64&  12216& 580&50 Cas&A2V             &   9553&  3.90&  2.6&    88& +0.15 & +0.40 &  28 & +0.21 & 1172 & 1.05 &       \\
  65&  28355&1414&79 Tau&A7V             &   7809&  3.98&  4.0&    87& +0.15 & $-$0.35 & 115 & $-$0.42 & 2557 & 1.00 &Am    \\
  66& 222345&8968&$\omega^{1}$ Aqr&A7IV  &   7487&  3.88&  3.8&    86& $-$0.07 & +0.03 & 168 & $-$0.37 & 3428 & 1.26 &       \\
  67&  74198&3449&$\gamma$ Cnc&A1IV      &   9381&  4.11&  2.8&    85& +0.26 & $-$0.14 &  18 & $-$0.20 &  931 & 1.01 &       \\
  68&  27934&1387&$\kappa^{1}$ Tau&A7IV-V&   8159&  3.84&  4.0&    83& +0.00 & $-$0.05 & 116 & $-$0.12 & 2817 & 0.85 &       \\
  69&  25490&1251&$\nu$ Tau&A1V          &   9077&  3.93&  3.2&    82& +0.01 & $-$0.44 &  15 & $-$0.41 &  934 & 0.98 &       \\
  70&  29388&1473&90 Tau&A6V             &   8194&  3.88&  4.0&    82& $-$0.03 & $-$0.03 & 116 & $-$0.10 & 2795 & 0.92 &       \\
  71&  79469&3665&$\theta$ Hya&B9.5V     &  10510&  4.20&  1.4&    82& $-$0.01 & $\cdots$ &  $\cdots$& $-$0.37 &  308 & 1.18 &       \\
  72&  28226&1403&         &Am           &   7361&  4.01&  3.6&    81& +0.25 & $-$0.38 & 128 & $-$0.60 & 2681 & 1.02 &Am    \\
  73& 207098&8322&$\delta$ Cap&A5mF2 (IV)&   7312&  4.06&  3.6&    81& +0.10 & $-$0.45 & 123 & $-$0.74 & 2325 & 1.18 &Am    \\
  74& 216627&8709&$\delta$ Aqr&A3V       &   8587&  3.59&  3.7&    79& $-$0.26 & $-$0.06 &  63 & $-$0.20 & 1750 & 1.02 &       \\
  75&  33641&1689&$\mu$ Aur&A4m          &   7961&  4.21&  4.0&    79& +0.08 & $-$0.56 &  85 & $-$0.58 & 1910 & 1.15 &Am    \\
  76&  12111& 575&48 Cas&A3IV            &   7910&  4.08&  4.0&    76& $-$0.26 & $-$0.33 & 112 & $-$0.55 & 2062 & 1.04 &       \\
  77&  23281&1139&         &A5m          &   7761&  4.19&  4.0&    76& +0.03 & $-$0.44 & 112 & ($-$1.21) & (1106) & 1.85 &Am, WBK\\
  78& 192640&7736&29 Cyg&A2V             &   8845&  3.86&  3.5&    74& $-$1.54 & $-$0.41 &  25 & $-$0.60 &  915 & 1.03 &       \\
  79&  11636& 553&$\beta$ Ari&A5V...     &   8294&  4.12&  3.9&    73& +0.10 & $-$0.18 &  96 & $-$0.38 & 1885 & 0.97 &Am    \\
  80& 173880&7069&111 Her&A5III          &   8567&  4.27&  3.8&    72& +0.20 & $-$0.10 &  82 & $-$0.19 & 1908 & 1.01 &       \\
  81&   5448& 269&$\mu$ And&A5V          &   8147&  3.82&  4.0&    72& $-$0.14 & $-$0.14 & 107 & $-$0.27 & 2383 & 0.87 &       \\
  82&  17093& 812&38 Ari&A7III-IV        &   7541&  3.95&  3.8&    69& $-$0.25 & $-$0.20 & 144 & $-$0.39 & 3191 & 1.15 &       \\
  83&  28319&1412&$\theta^{2}$ Tau&A7III &   7789&  3.68&  4.0&    68& $-$0.11 & $-$0.15 & 134 & $-$0.20 & 3403 & 1.35 &       \\
  84&  95382&4294&59 Leo&A5III           &   8017&  3.95&  4.0&    68& $-$0.10 & $-$0.08 & 128 & $-$0.22 & 2785 & 1.00 &       \\
  85& 140436&5849&$\gamma$ CrB&A1Vs      &   9274&  3.89&  3.0&    68& $-$0.21 & +0.05 &  25 & ($-$0.42) &  (762) & 1.98 &WBK  \\
  86&  20320& 984&$\zeta$ Eri&A5m        &   7505&  3.91&  3.8&    67& $-$0.13 & $-$0.76 &  90 & $-$1.27 & 1228 & 1.42 &Am    \\
  87&  13161& 622&$\beta$ Tri&A5III      &   7957&  3.68&  4.0&    65& $-$0.40 & $-$0.44 &  90 & $-$0.40 & 2408 & 1.35 &       \\
  88&  27045&1329&$\omega^{2}$ Tau&A3m   &   7552&  4.26&  3.8&    62& +0.32 & $-$0.45 & 117 & $-$0.80 & 1892 & 1.33 &Am    \\
  89& 200499&8060&$\eta^{2}$ Cap&A5V     &   8081&  3.95&  4.0&    61& $-$0.21 & $-$0.10 & 120 & $-$0.49 & 1937 & 1.13 &       \\
  90&  29499&1480&         &A5m          &   7638&  4.08&  3.9&    61& +0.32 & +0.16 & 175 & ($-$0.86) & (1722) & 3.42 &Am, WBK\\
  91& 116656&5054&$\zeta$ UMa&A2V        &   9317&  4.10&  2.9&    59& +0.27 & $-$0.09 &  22 & +0.00 & 1215 & 1.09 &       \\
  92& 198639&7984&56 Cyg&A4me...         &   7921&  4.09&  4.0&    59& +0.03 & $-$0.22 & 123 & $-$0.32 & 2649 & 1.03 &       \\
  93& 130841&5531&$\alpha^{2}$ Lib&A3IV  &   8079&  3.96&  4.0&    58& $-$0.37 & $-$1.12 &  27 & ($-$0.69) & (1554) & 2.05 &Am, WBK\\
  94&  30121&1511& 4    Cam&A3m          &   7700&  3.98&  3.9&    57& +0.25 & $-$0.58 &  98 & $-$0.98 & 1469 & 1.33 &Am    \\
  95&  30210&1519&         &Am...        &   7927&  3.94&  4.0&    56& +0.44 & $-$0.62 &  77 & $-$1.36 &  840 & 1.46 &Am    \\
  96&  29479&1478&$\sigma^{1}$ Tau&A4m   &   8406&  4.14&  3.9&    56& +0.35 & $-$0.38 &  66 & $-$0.47 & 1559 & 0.96 &Am    \\
  97& 222603&8984&$\lambda$ Psc&A7V      &   7757&  3.99&  4.0&    55& $-$0.19 & $-$0.13 & 143 & $-$0.34 & 2911 & 1.14 &       \\
  98& 212061&8518&$\gamma$ Aqr&A0V       &  10384&  3.95&  1.5&    54& $-$0.01 & $\cdots$ &  $\cdots$& $-$0.06 &  422 & 1.21 &       \\
  99&  89021&4033&$\lambda$ UMa&A2IV     &   8861&  3.61&  3.5&    52& +0.06 & $-$0.24 &  28 & $-$0.29 & 1230 & 0.97 &       \\
 100& 195725&7850&$\theta$ Cep&A7III     &   7816&  3.74&  4.0&    49& +0.14 & $-$0.17 & 131 & $-$0.43 & 2569 & 0.99 &Am    \\
 101&  43378&2238&2 Lyn&A2Vs             &   9210&  4.09&  3.0&    46& $-$0.15 & $-$0.07 &  28 & $-$0.10 & 1191 & 0.89 &       \\
 102&  27819&1380&$\delta^{2}$ Tau&A7V   &   8047&  3.95&  4.0&    45& $-$0.04 & $-$0.01 & 132 & $-$0.14 & 2977 & 0.93 &       \\
 103&  95418&4295&$\beta$ UMa&A1V        &   9489&  3.85&  2.7&    44& +0.21 & $-$0.14 &  10 & $-$0.08 &  891 & 1.00 &       \\
 104& 218396&8799&         &A5V          &   7091&  4.06&  3.3&    41& $-$0.58 & $-$0.74 & 101 & ($-$0.89) & (2255) & 1.72 &WBK  \\
 105&  84107&3861&15 Leo&A2IV            &   8665&  4.31&  3.7&    38& +0.01 & $-$0.36 &  50 & $-$0.37 & 1438 & 0.95 &       \\
 106& 204188&8210&         &A8m          &   7622&  4.21&  3.9&    36& +0.02 & $-$0.38 & 123 & $-$0.51 & 2577 & 1.13 &Am    \\
 107&  33204&1670&         &A5m          &   7530&  4.06&  3.8&    34& +0.28 & $-$0.32 & 133 & ($-$1.52) &  (902) & 3.06 &Am, WBK\\
 108& 141795&5892&$\epsilon$ Ser&A2m     &   8367&  4.24&  3.9&    32& +0.23 & $-$0.66 &  46 & $-$0.70 & 1264 & 1.01 &Am    \\
 109& 173648&7056&$\zeta^{1}$ Lyr&Am     &   8004&  3.90&  4.0&    32& +0.29 & $-$0.38 &  96 & $-$0.50 & 2036 & 0.95 &Am    \\
 110& 112185&4905&$\epsilon$ UMa&A0p     &   9407&  3.61&  2.8&    32& +0.17 & $\cdots$ &  $\cdots$& $-$1.46 &  271 & 1.18 &       \\
 111&  27628&1368&60 Tau&A3m             &   7218&  4.05&  3.5&    30& +0.32 & $-$0.67 & 104 & ($-$2.57) &  (379) & 3.21 &Am, WBK\\
 112&  28546&1428&81 Tau&Am              &   7640&  4.17&  3.9&    28& +0.22 & $-$0.39 & 122 & $-$0.62 & 2249 & 0.99 &Am    \\
 113& 172167&7001&$\alpha$ Lyr&A0Vvar    &   9435&  3.99&  2.7&    21& $-$0.55 & $-$0.47 &   7 & $-$0.62 &  546 & 1.00 &       \\
 114&  60179&2891&$\alpha$ Gem&A2Vm      &   9122&  3.88&  3.2&    19& $-$0.03 & $-$0.74 &   7 & $-$0.68 &  660 & 1.21 &Am    \\
 115&  95608&4300&60 Leo&A1m             &   8972&  4.20&  3.3&    18& +0.30 & $-$0.83 &  11 & $-$0.73 &  754 & 1.01 &Am    \\
 116&  48915&2491&$\alpha$ CMa&A0m...    &   9938&  4.31&  2.1&    16& +0.42 & $-$0.52 &   3 & $-$0.39 &  502 & 1.11 &Am    \\
 117&  27749&1376&63 Tau&A1m             &   7448&  4.21&  3.7&    13& +0.47 & $-$1.16 &  51 & ($-$2.18) &  (487) & 2.24 &Am, WBK\\
 118&  33254&1672&16 Ori&A2m             &   7747&  4.14&  3.9&    13& +0.38 & $-$0.95 &  59 & ($-$1.75) &  (623) & 3.22 &Am, WBK\\
 119&  72037&3354& 2 UMa&A2m             &   7918&  4.16&  4.0&    12& +0.21 & $-$1.03 &  43 & $-$1.10 & 1121 & 1.12 &Am    \\
 120&  27962&1389&$\delta^{3}$ Tau&A2IV  &   8923&  3.94&  3.4&    11& +0.23 & $-$0.24 &  31 & $-$0.32 & 1174 & 1.15 &Am    \\
 121&  47105&2421&$\gamma$ Gem&A0IV      &   9115&  3.49&  3.2&    11& $-$0.06 & $-$0.03 &  21 & $-$0.22 & 1002 & 1.47 &       \\
 122&  40932&2124&$\mu$ Ori&Am...        &   8005&  3.93&  4.0&    11& $-$0.12 & $-$0.76 &  58 & $-$0.60 & 1827 & 1.31 &Am    \\
\end{longtable}

\setcounter{table}{1}
\begin{table}[h]
\begin{minipage}{180mm}
\scriptsize
\caption{Atomic data of Ca~{\sc i} 6162 and Ca~{\sc ii} 3934 lines used in 
the spectrum synthesis.}
\begin{center}
\begin{tabular}
{cccccccc}\hline \hline
RMT & $\lambda$ & $\chi_{\rm low}$ & $\log gf$ & Gammar & Gammas &
Gammaw & Source \\
   &   (\AA)  & (eV)  & (dex)  & (dex)  & (dex) & (dex) &  \\
\hline
1 & 3933.655 &   0.000 & $-$2.623 &   8.21 & $-$5.73 & $-$7.76 & VALD \\
1 & 3933.657 &   0.000 & $-$4.293 &   8.21 & $-$5.73 & $-$7.76 & VALD \\
1 & 3933.659 &   0.000 & $-$1.576 &   8.21 & $-$5.73 & $-$7.76 & VALD \\
1 & 3933.660 &   0.000 & $-$2.765 &   8.21 & $-$5.73 & $-$7.76 & VALD \\
1 & 3933.661 &   0.000 & $-$2.084 &   8.21 & $-$5.73 & $-$7.76 & VALD \\
1 & 3933.664 &   0.000 &   +0.092 &   8.21 & $-$5.73 & $-$7.76 & VALD \\
\hline
3 & 6162.173 &   1.899 & +0.100 &   7.82 & $-$5.07 & $-$7.59 & KB   \\
\hline
\end{tabular}
\end{center}
\scriptsize
Note. \\
The first four columns are self-explanatory.
The damping parameters are given in the following three columns:\\
Gammar is the radiation damping width (s$^{-1}$), $\log\gamma_{\rm rad}$.\\
Gammas is the Stark damping width (s$^{-1}$) per electron density (cm$^{-3}$) 
at $10^{4}$ K, $\log(\gamma_{\rm e}/N_{\rm e})$.\\
Gammaw is the van der Waals damping width (s$^{-1}$) per hydrogen density 
(cm$^{-3}$) at $10^{4}$ K, $\log(\gamma_{\rm w}/N_{\rm H})$. \\
KB $\cdots$ Kurucz and Bell's (1995) compilation, VALD $\cdots$ VALD database
(Ryabchikova et al. 2015)\\
Note that Ca~{\sc ii} 3934 line was treated as a single line
(with the $gf$-weighted mean parameters of $\lambda$ = 3933.664~\AA\ and 
$\log gf = +0.105$) in the evaluation of equivalent width by using Kurucz's (1993)
WIDTH9 program, since the differences between the components (corresponding to
different isotopes) are negligible ($\ltsim 0.01$~\AA) compared to
the intrinsic stellar line width.  
\end{minipage}
\end{table}

\section{Spectrum fitting analysis}

\subsection{Outline of the method}

The procedures of abundance determination are essentially the same as in Paper~I
(cf. section~4 therein). First, by applying the numerical algorithm described 
in Takeda (1995), the best fit between the synthetic and observed spectra was accomplished 
in the specified wavelength region while varying the abundances of important elements 
($A_{1}$, $A_{2}$, $\ldots$)\footnote{Abundances of the remaining elements were 
fixed.}, $v \sin i$ (projected rotational velocity), and $\Delta \lambda$ 
(radial velocity or wavelength shift). 
Then, the equivalent width ($W$) of the relevant line of interest (Ca in the present 
case) was ``inversely'' evaluated from the best-fit abundance solution with 
the same atmospheric model/parameters as used in the spectrum-fitting analysis. 
This $W$ was further used to estimate the abundance uncertainties 
due to ambiguities of atmospheric parameters by perturbing
the standard values interchangeably. Theoretical line-formation calculations 
(necessary for spectrum synthesis or for evaluation of $W$) were made under the assumption 
of LTE (Local Thermodynamic Equilibrium) throughout this study, since the non-LTE
effect is likely to be insignificant for the relevant Ca lines (see appendix~1).

\subsection{Analysis of Ca~I 6162 and Ca~{\sc ii} 3934 lines}

This spectrum fitting for the 6140--6170~\AA\ region comprising Ca~{\sc i} lines
was already done in Paper~I, where the atomic parameters of all spectral lines 
were taken from Kurucz and Bell's (1995) compilation and the abundances of five 
elements (O, Si, Ca, Fe, and Ba) were varied. Hereinafter, the quantities derived 
from the fitting analysis in this wavelength region are denoted with suffix ``61''.
How the theoretical spectrum for the converged parameters fit the observed spectrum 
is demonstrated in figure~2 of Paper~I, and the resulting $A_{61}$(Ca) and 
$v\sin i_{61}$ for each star are presented in table~1,
Among the several Ca~{\sc i} lines of appreciable strengths included in 
this 6140--6170~\AA\ region, the main focus is placed on the strongest 
Ca~{\sc i} 6162.17 line (multiplet~3, $\chi_{\rm low} = 1.90$~eV; see 
table~2 for the atomic parameters of this line). Therefore, the equivalent widths 
of this Ca~{\sc i}~6162 line ($W_{61}$) were calculated inversely from 
$A_{61}$(Ca), which are also given in table~1.

Similarly, in order for the main purpose of determining the Ca abundance 
from the strong Ca~{\sc ii} 3934 line, synthetic spectrum fitting analysis was 
done in the same manner for the 3910--3955~\AA\ region, where the atomic parameters
were taken from the VALD database (Ryabchikova et al. 2015) and the abundances of
Ca and Fe were varied. The quantities derived from this 3910--3955~\AA\ fitting 
are denoted with suffix ``39''.
The resulting solutions of $A_{39}$(Ca) and $v\sin i_{39}$ for each star are 
given in table~1, while the theoretical spectrum for the converged parameters
is compared with the observed spectrum is displayed for each star in figure~1.
Likewise, the equivalent widths ($W_{39}$) of the Ca~{\sc ii} 3933.68 line 
(multiplet~1, $\chi_{\rm low} = 0.00$~eV; see table~2 for the adopted atomic 
parameters) were evaluated inversely from $A_{39}$(Ca) and are listed in table~1.

The equivalent widths, Ca abundances, and their sensitivities
to perturbations of atmospheric parameters ($T_{\rm eff}$, $\log g$, 
and $v_{\rm t}$) estimated in the same manner as in Paper~I 
(see the caption in figure~2 of this paper) are plotted against 
$T_{\rm eff}$ in figure~2 (Ca~{\sc i} 6162) and figure~3 (Ca~{\sc ii} 3934).

\setcounter{figure}{0}
\begin{figure*}
  \begin{center}
    \FigureFile(150mm,200mm){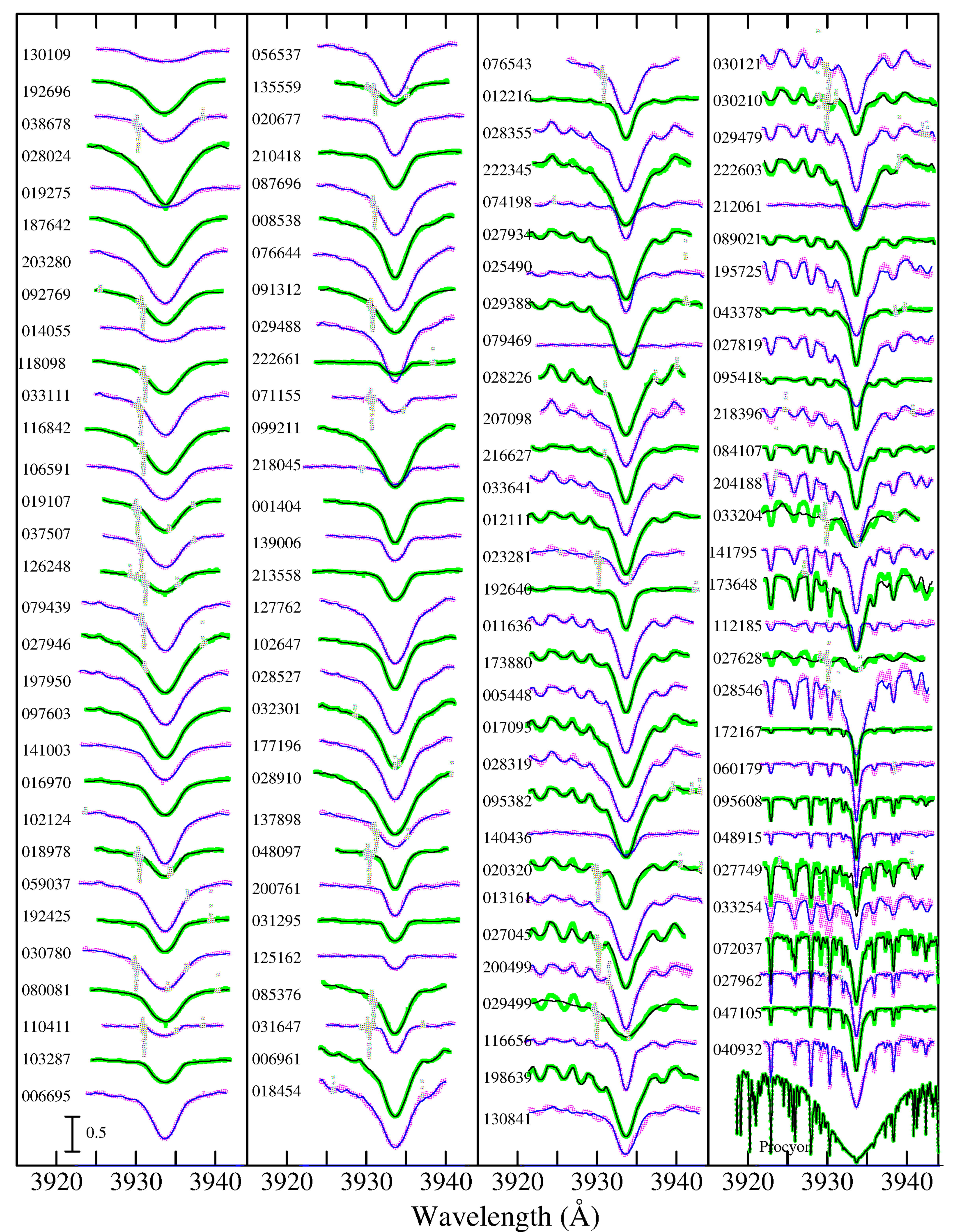}
  \end{center}
\caption{Synthetic spectrum fitting at the 3910--3955~\AA\ region
(comprising the Ca~{\sc ii} 3934 line) accomplished by varying the 
projected rotational velocity ($v_{\rm e} \sin i_{39}$) along with 
the abundances of Ca and Fe. 
The best-fit theoretical spectra are shown by black/blue solid lines, 
while the observed data are plotted by green/pink symbols (note that 
the masked portions of the observed spectra which were discarded 
because of the spectrum defect are indicated by gray symbols).
In each panel, the spectra are arranged in the descending order 
of $v_{\rm e} \sin i_{61}$ as in table 1, and an offset of 0.5 is 
applied to each spectrum (indicated by the HD number, where that 
of ``Weak Broad K-line'' star is colored in red) 
relative to the adjacent one. The case of Procyon (standard star) is 
displayed at the bottom of the rightmost panel.
}
\end{figure*}

\setcounter{figure}{1}
\begin{figure}
  \begin{center}
    \FigureFile(70mm,70mm){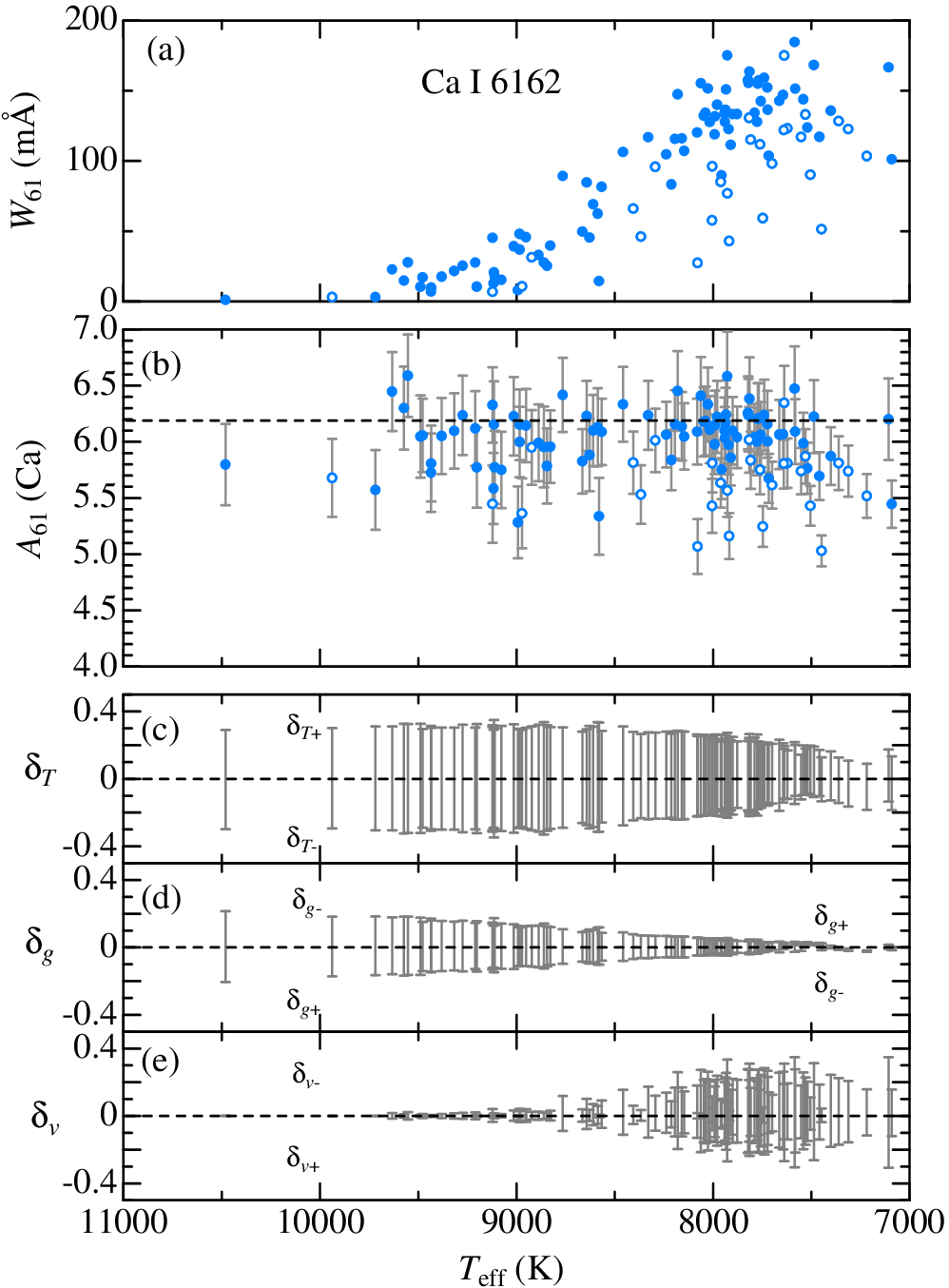}
  \end{center}
\caption{
Abundance-related results derived from 6140--6170~\AA\ fitting 
(comprising Ca~{\sc i} lines) plotted against $T_{\rm eff}$. 
(a) Equivalent width of Ca~{\sc i} 6162 line ($W_{61}$), (b) 
$A_{61}$(Ca) (logarithmic number abundance of Ca in the usual normalization 
of $A_{\rm H} = 12.00$), (c) $\delta_{T+}$ and $\delta_{T-}$ (Ca abundance 
variations in response to $T_{\rm eff}$ changes of
+300~K and $-300$~K), (d) $\delta_{g+}$ and $\delta_{g-}$ 
(Ca abundance variations in response to $\log g$ changes 
of $+0.3$~dex and $-0.3$~dex), and (e) $\delta_{v+}$ and 
$\delta_{v-}$ (abundance variations in response to 
changing $v_{\rm t}$ as $v_{\rm t} \times 1.3$ and $v_{\rm t} / 1.3$).
In panels (a) and (b), Am stars (see the remark in table~1) are expressed
in open symbols. The error bars attached in the symbols in panel (b) are
$\delta_{Tgv} \equiv \sqrt{\delta_{T}^{2} + \delta_{g}^{2} + \delta_{v}^{2}}$,
where $\delta_{T} \equiv (|\delta_{T+}|+|\delta_{T-}|)/2$,
$\delta_{g} \equiv (|\delta_{g+}|+|\delta_{g-}|)/2$, and 
$\delta_{v} \equiv (|\delta_{v+}|+|\delta_{v-}|)/2$.  
}
\end{figure}

\setcounter{figure}{2}
\begin{figure}
  \begin{center}
    \FigureFile(70mm,70mm){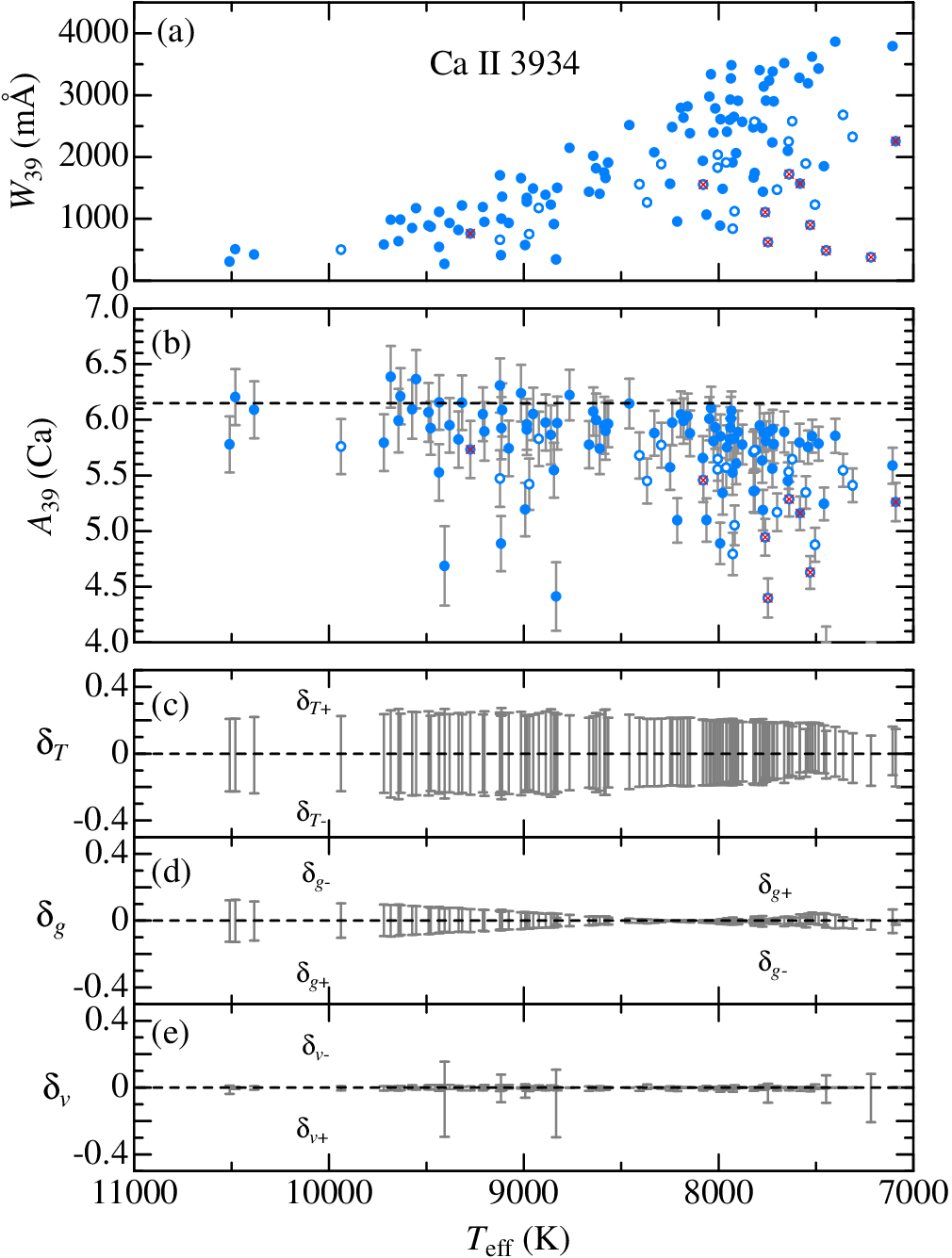}
  \end{center}
\caption{
Abundance-related results derived from 3910--3955~\AA\ fitting 
(comprising Ca~{\sc ii} 3934) plotted against $T_{\rm eff}$. 
(a) Equivalent width of Ca~{\sc ii} 3934 line ($W_{39}$), (b) $A_{39}$(Ca) 
(Ca abundance). Panels (c), (d), and (e) show the abundance perturbations
in response to changing $T_{\rm eff}$, $\log g$, and $v_{\rm t}$ evaluated
in the same manner as in figure~2.
Regarding the symbols in panels (a) and (b), Am stars are distinguished by
open symbols as in figure~2, while overplotted red crosses indicate 
``broad weak K line'' stars (labeled as BWK in the remark of table~1).   
}
\end{figure}

\section{Discussion}

\subsection{Trend of Ca line strengths}

It may be worthwhile to examine the behaviors of the Ca~{\sc i} 6162 and 
Ca~{\sc ii} 3934 lines from the viewpoint of the parameter dependence.
As can be seen from figure~2a and figure~3a, $W_{61}$ as well as $W_{39}$
are progressively weakened with an increase in $T_{\rm eff}$, the gradient 
being especially steeper for the former. This may be interpreted as 
the $T_{\rm eff}$-dependence of the number population of the lower level
from which the line forms.

According to the Saha--Boltzmann equation, the $T$-dependence for the number 
population ($n_{i}^{m}$) of level $i$ ($i=0$ for the ground level) at 
ionization stage $m$ (=I, II, III) can be expressed as
\begin{equation}
n_{i}^{m} \propto n_{0}^{m+1} \exp{[(\chi_{\rm ION}^{m}-\chi_{i}^{m})/(kT)]} 
\end{equation}
and
\begin{equation}
n_{i}^{m} \propto n_{0}^{m} \exp{[-\chi_{i}^{m}/(kT)]}
\end{equation}
where $\chi_{i}^{m}$ is the excitation potential of level $i$ at ionization stage $m$,
$\chi_{\rm ION}^{m}$ is the ionization potential (from the ground level of
ionization stage $m$), and $k$ is the Boltzmann constant (note that the $T$-dependent 
term proportional to $\propto T^{-3/2}$ was neglected in equation (1) because
it is insignificant compared to the exponential term).
Then, $n_{i}^{\rm I}$ and $n_{i}^{\rm II}$ can be written in terms of $n_{0}^{\rm III}$ as 
\begin{equation}
n_{i}^{\rm I} \propto n_{0}^{\rm III} 
  \exp{[(\chi_{\rm ION}^{\rm I} + \chi_{\rm ION}^{\rm II} - \chi_{i}^{\rm I})/(kT)]}
\end{equation}
and
\begin{equation}
n_{i}^{\rm II} \propto n_{0}^{\rm III} \exp{[(\chi_{\rm ION}^{\rm II}-\chi_{i}^{\rm II})/(kT)]}. 
\end{equation}
Here, the important point is that Ca is mainly in the twice-ionized stage (Ca~{\sc iii}) 
in the photosphere ($\tau_{5000} \sim 1$) of A-type stars (cf. figure~4), which
means that $n_{0}^{\rm III}$ (dominant stage) is not much sensitive to $T$. 

Accordingly, the number populations of the lower level for the Ca~{\sc i} 6162 and 
Ca~{\sc ii} 3934 lines ($n_{\rm low,61}$, $n_{\rm low,39}$) have the $T$-dependence of
\begin{equation}
n_{\rm low,61} \propto \exp{[(6.11 + 11.87 - 1.90)/(kT)]}
\end{equation}
and
\begin{equation}
n_{\rm low,39} \propto \exp{[(11.87 - 0.00)/(kT)]},
\end{equation}
where $\chi_{\rm low,61}^{\rm I} = 1.90$~eV (for Ca~{\sc i} 6162), 
$\chi_{\rm low,39}^{\rm II} = 0.00$~eV (for Ca~{\sc ii} 3934), 
$\chi_{\rm ION}^{\rm I} = 6.11$~eV,
and $\chi_{\rm ION}^{\rm II} = 11.87$~eV.
Equations (5) and (6) explain the reason why $W_{61}$ fades out more quickly 
(compared to $W_{39}$) with an increase in $T_{\rm eff}$.

Since the abscissa of curve of growth is written as 
$A + \log n_{\rm low} + \log gf$ (+const.), empirical curves of growth
in the present cases of two lines can be constructed based on 
the $W$ and $A$ results for 122 stars given in table~1 by plotting 
the $\log W_{61}$ vs. $A_{61}+ \log n_{\rm low,61}$ (for Ca~{\sc i} 6162) 
and $\log W_{39}$ vs. $A_{39}+ \log n_{\rm low,39}$ (for Ca~{\sc ii} 3934) 
relations as shown in figures~5a and 5b, respectively, 
where $T_{\rm eff}$ has been assigned to $T$.
It can be seen from figure~5 that Ca~{\sc i} 6162 line is on the linear 
($W \propto A$) through flat part of the curve of growth, while Ca~{\sc ii} 3934 
line is mostly on the damping part of the curve of growth showing the square-root 
dependence of $W$ upon $A$ (except for several lines showing anomalous weakness).

\setcounter{figure}{3}
\begin{figure}
  \begin{center}
    \FigureFile(70mm,70mm){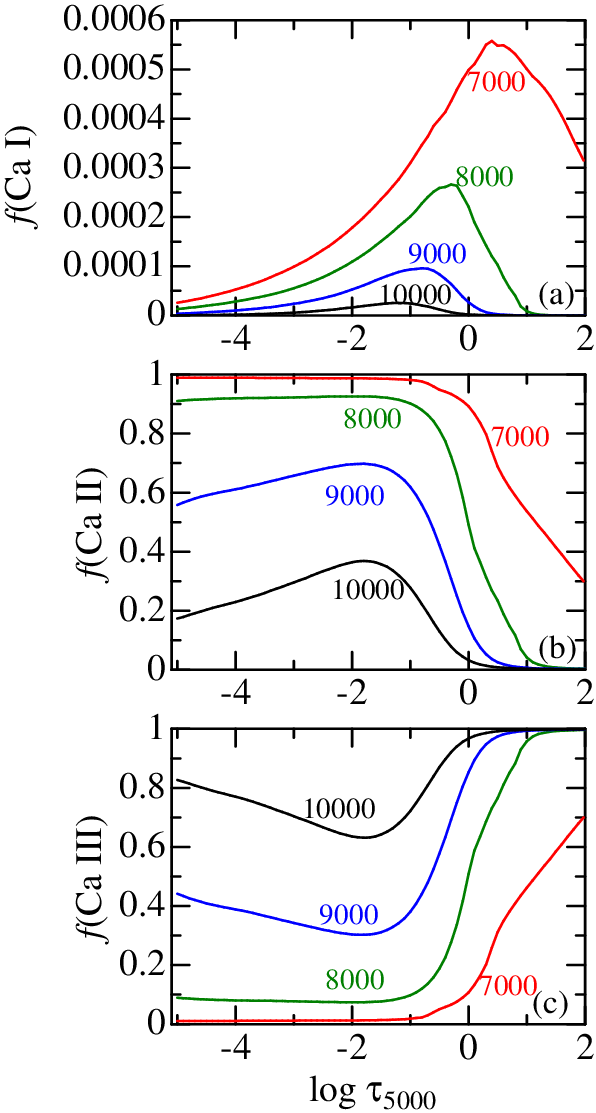}
  \end{center}
\caption{
Number population fraction ($f$) of (a) neutral, (b) once-ionized, 
and (c) twice-ionized calcium species relative to the total Ca atoms
[e.g., $f$(Ca~{\sc i}) $\equiv N$(Ca~{\sc i})/$N_{\rm total}^{\rm Ca}$],
plotted against the continuum optical depth at 5000~\AA. 
Calculations were done for four $\log g = 4.0$ model of different
$T_{\rm eff}$ (7000, 8000, 9000, and 10000~K) as indicated in each panel.
}
\end{figure}

\setcounter{figure}{4}
\begin{figure}
  \begin{center}
    \FigureFile(80mm,50mm){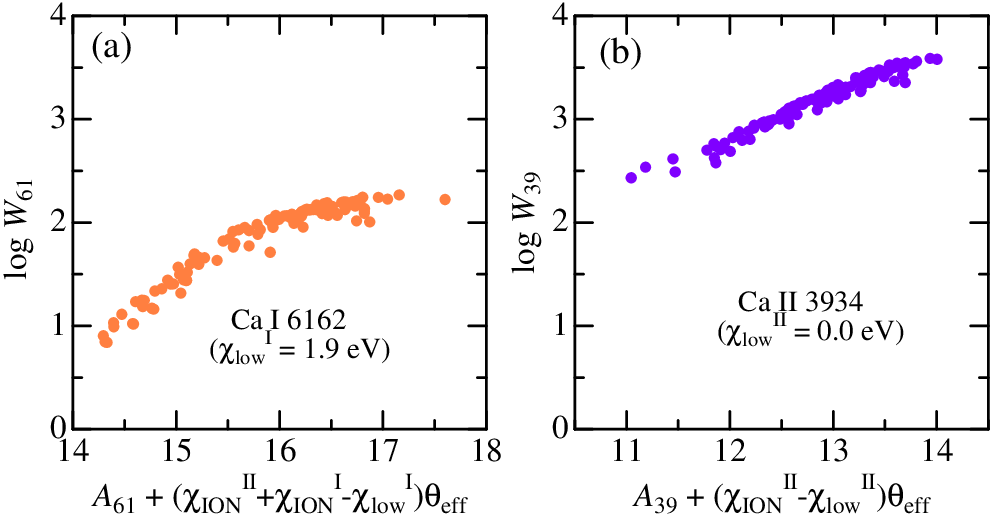}
  \end{center}
\caption{
Curves of growth for the Ca~{\sc i} 6162 and Ca~{\sc ii} 3934 lines 
constructed from the equivalent widths ($W$) and the corresponding abundances 
($A$) given in table~1. 
In panel (a) are plotted the $\log W_{61}$ values against  
$A_{61} + (\chi_{\rm ION}^{\rm II} + \chi_{\rm ION}^{\rm I} - \chi_{\rm low,61}^{\rm I})\theta_{\rm eff}$.
Here, $\chi_{\rm ION}^{\rm II}$ (=~11.87~eV) and $\chi_{\rm ION}^{\rm I}$ (=~6.11~eV) are 
the ionization potentials of Ca~{\sc ii} and Ca~{\sc i}, respectively, and
$\chi_{\rm low,61}^{\rm I}$ (=~1.9~eV) is the lower excitation potential of 
Ca~{\sc i} 6162, and $\theta_{\rm eff} \equiv 5040/T_{\rm eff}$. 
Similarly, panel (b) shows the relations between $\log W_{39}$ and 
$A_{39} + (\chi_{\rm ION}^{\rm II} - \chi_{\rm low,39}^{\rm II})\theta_{\rm eff}$,
where $\chi_{\rm low,39}^{\rm II}$ (=~0.0~eV) is the lower excitation potential of 
Ca~{\sc ii} 3934.
}
\end{figure}

\subsection{Abundance sensitivity to atmospheric parameters}

Some comments may be appropriate on the abundance sensitivities due 
to uncertainties of atmospheric parameters (cf. panels (c)--(e) in figures 2 and 3)  
based on what has been described in subsection~5.1.

As seen from these figure panels, Ca abundances are most sensitive to 
a change in $T_{\rm eff}$ (especially for early A stars).
The strong $T_{\rm eff}$-sensitivity can be explained by equations (5) and (6), 
which suggest $|\Delta A_{61}| \sim 16.08\times(5040/T_{\rm eff}^{2})|\Delta T_{\rm eff}|$
and 
$|\Delta A_{39}| \sim 11.87\times(5040/T_{\rm eff}^{2})|\Delta T_{\rm eff}|$.
Inserting $T_{\rm eff} = 8000$~K and $\Delta T_{\rm eff} = 300$~K, for example, 
we have $|\Delta A_{61}| \sim 0.4$~dex and $|\Delta A_{39}| \sim 0.3$~dex,
consistently with figures 2c and 3c.

In contrast, Ca abundances are not so sensitive to a change of $\log g$ (cf. 
figures 2d and 3d), especially for the damping-dominated strong Ca~{\sc ii} 3934 
line, because two effects caused by an increase $\log g$ compensate with each other: 
i.e., (i) enhanced continuum opacity (line weakening) and (ii) growth of damping 
wing (line strengthening). 

Regarding the effect of changing $v_{\rm t}$ on the resulting abundances,
the situation is markedly different between both lines (cf. figures 2e and 3e).
Since the strength of Ca~{\sc i} 6162 ($W_{61}$) widely varies from very weak (linear 
part) at higher $T_{\rm eff}$ to strongly saturated (flat part) at lower $T_{\rm eff}$, 
$A_{61}$ is almost $v_{\rm t}$-independent for early-A stars, while it is very 
sensitive to a choice of $v_{\rm t}$ for late-A stars ($T_{\rm eff} \ltsim 8000$~K). 
In contrast, the role of $v_{\rm t}$ in determining $A_{39}$ is essentially 
insignificant, because Ca~{\sc ii} 3934 line is too strong (damping part) to be
affected by $v_{\rm t}$, though some exceptional $v_{\rm t}$-sensitive cases exist 
where $W_{39}$ is as low as several hundred m\AA. 

\subsection{Stars with anomalous Ca~II 3934 line profile}

Unexpectedly, embarrassing results were noticed from the 3910--3955~\AA\ region 
fitting analysis (cf. section 4). That is, $v\sin i_{39}$ values for some of 
the program stars were found to be considerably discrepant from the $v\sin i_{61}$ 
solutions (regarded as standard $v\sin i$) derived from the 6140--6170~\AA\ region 
fitting in Paper~I, despite that both are based on the same spectra. 
These two $v\sin i$ values are compared in figures~6a and 6b, which 
elucidate that, in an appreciable fraction of slow rotators 
($v \sin i_{61} \ltsim 100$~km~s$^{-1}$), $v \sin i_{39}$ is considerably 
larger than $v \sin i_{61}$ even by up to $\sim$~200--300\% in extreme cases.
It is on the side of $v \sin i_{39}$ (not $v\sin i_{61}$) that must be wrong
in this context, because $v \sin i_{61}$ and $v \sin i_{58}$ (determined in Paper~I 
from the spectrum fitting in the 5880--5905~\AA\ region comprising Na~{\sc i} 
5890/5896 lines) are still consistent with each other as shown in figures~6c and 6d.

A close inspection of these outliers revealed that the cause of this
discrepancy stems from the anomalous profile of the Ca~{\sc ii} 3934 line;
i.e., it is unusually broad for its weak strength.
The automatic fitting algorithm adopted in this study accomplishes the best 
parameter solutions that minimize the difference ($\chi^{2}$) between 
the theoretical and observed spectra in the relevant wavelength region.
As the Ca~{\sc ii} K line is the dominant spectral feature in 3910--3955~\AA,
this program tries to adjust $v\sin i$ so as fit this broad Ca~{\sc ii} 3934 line, 
which eventually resulted in an spuriously large $v\sin i_{39}$.
This is reason why an appreciable inequality ($v \sin i_{39} > v \sin i_{61}$)
is seen in these stars. 

Since such obtained $v\sin i_{39}$ solution is an appreciable overestimation
for the actual projected rotational velocity, metallic line features other than 
Ca~{\sc ii} 3934 can not be reproduced any more, which results in considerably
poor fit for such lines (e.g., Fe~{\sc i} lines); see, e.g., the spectrum fit
appearance in figure~1 for those peculiar objects such as HD~29499 (\#90), 
HD~33204 (\#107), HD~27628 (\#111), and HD~33254 (\#118).
On the other hand, if the true $v \sin i$ is used as fixed, Ca~{\sc ii} K line 
profile can never be fitted no matter how $A_{39}$(Ca) is varied.
This situation is demonstrated in figure~7, where the observed spectrum of
HD~29499 (the star showing the largest anomaly of $v \sin i_{39}/ v \sin i_{61} = 3.4$)
is compared with theoretical spectra computed for 4 different Ca abundances but 
broadened with $v\sin i_{61} = 61$~km~s$^{-1}$ (real rotational velocity which 
adequately reproduces the profiles of neighboring metallic lines).
This figure clearly shows that the observed Ca~{\sc ii} K line is too broad 
(as if it is a damping-dominated line) despite that its strength is only mild 
(like lines in the flat-part of curve of growth).

Hereinafter, 10 especially anomalous stars with the criterion of 
$v \sin i_{39}/ v \sin i_{61} (\equiv \alpha) > 1.5$
(i.e., discrepancy greater than 50\%) are called ``Weak Broad K-line stars'' 
(abbreviated as WBK stars), which are marked by overplotted red 
crosses in figures~3a,b and figures~6a,b. They are also remarked in column 16 
of table~1 and their HD numbers inserted in figure~1 are colored in red.
The characteristics of these WBK stars can be summarized as follows.
\begin{itemize}
\item
They are mostly late A-type stars at $T_{\rm eff} \ltsim 8000$~K and show 
considerably low $A_{39}$ and small $W_{39}$ compared to other stars at 
the same $T_{\rm eff}$ (cf. figures~3a,b). 
\item
They are slow rotators of $v\sin i_{61} \ltsim 100$~km~s$^{-1}$ (cf. figures~6a,b).
\item
As can be expected from the two characteristics mentioned above, most of 
them are classified as Am, though this does not mean that Am stars 
and WBK stars are equivalent (Am stars which do not show WBK phenomenon
do exist).
\item
Since any satisfactory match could not be accomplished between the theoretical
and observed spectrum in the 3910--3955~\AA\ fitting, the resulting $A_{39}$
(or [Ca/H]$_{39}$) and $W_{39}$ for these stars (shown with parentheses in table~1) 
are unreliable\footnote{Since a number of weaker metallic lines in the neighborhood
of Ca~{\sc ii} 3934 are regarded as if they are noises in such a case of unattained fit, 
$A_{39}$ and $W_{39}$ values tend to be underestimated, because the continuum level
is wrongly placed too low.} 
and thus should be regarded with caution. 
\item
It is difficult to understand the physical mechanism causing such an extraordinary 
WBK phenomenon, which is apparently inexplicable within the framework of 
the classical line-formation theory. A possibility of Ca abundance stratification 
was examined (cf. appendix~2), which however turned out unsuccessful. 
\end{itemize}

\setcounter{figure}{5}
\begin{figure}
  \begin{center}
    \FigureFile(90mm,100mm){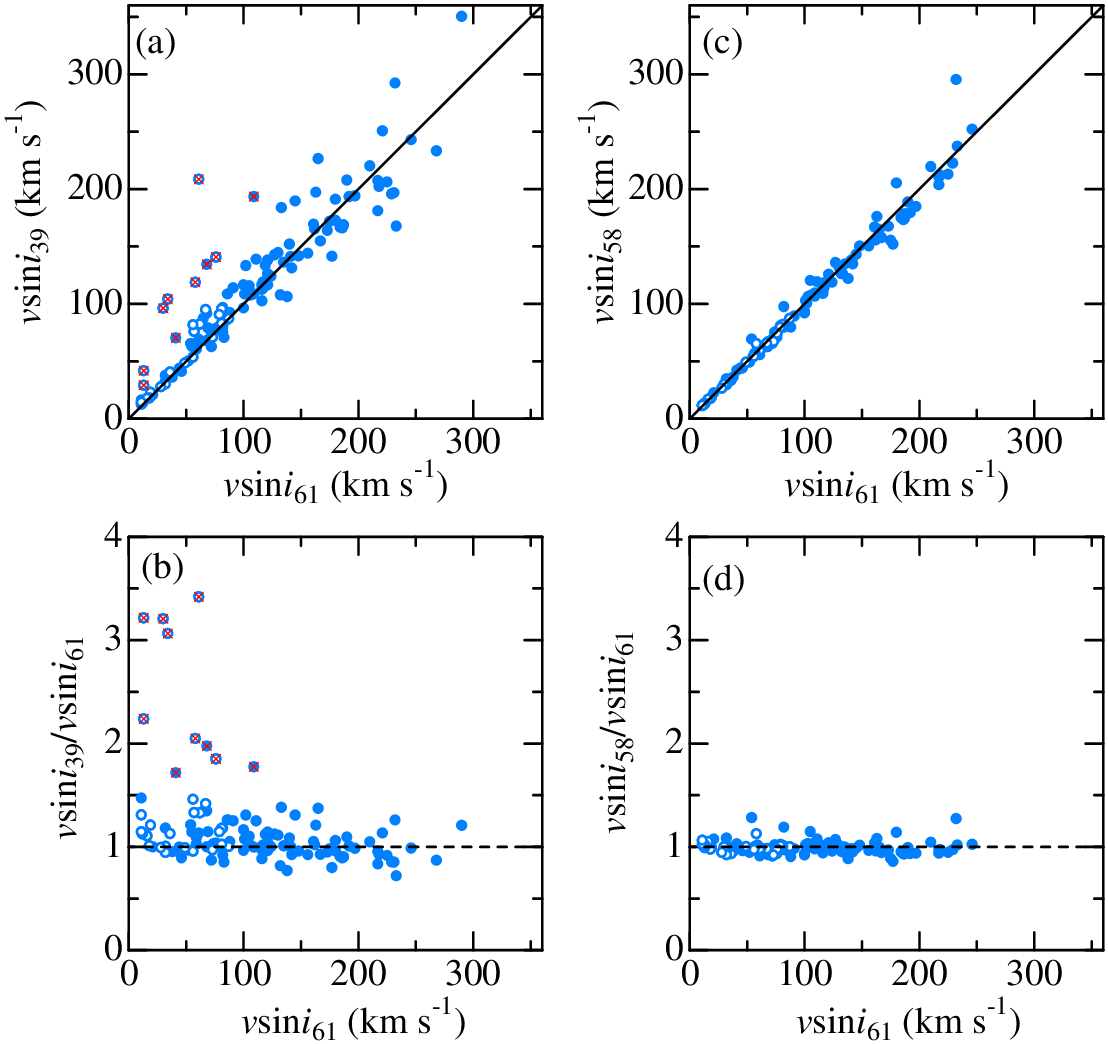}
  \end{center}
\caption{
Comparison of the standard $v\sin i$ values derived from 
6140--6170~\AA\ fitting in Paper~I ($v\sin i_{61}$)
with $v\sin i_{39}$ (from 3910--3955~\AA\ fitting in this study)
and $v\sin i_{58}$ (from 5880--5905~\AA\ fitting in Paper~I).
(a) $v\sin i_{39}$ vs. $v\sin i_{61}$, 
(b) $v\sin i_{39}/v\sin i_{61}$ vs. $v\sin i_{61}$,
(c) $v\sin i_{58}$ vs. $v\sin i_{61}$, and
(d) $v\sin i_{58}/v\sin i_{61}$ vs. $v\sin i_{61}$.
Am stars are distinguished by open symbols as in figures 2 and 3.
As done in figures~3a and 3b, 10 weak broad K-line stars ($v\sin i_{39}/v\sin i_{61} > 1.5$) 
are marked by overplotted red crosses in the left-hand panels (a) and (b), 
}
\end{figure}

\setcounter{figure}{6}
\begin{figure}
  \begin{center}
    \FigureFile(70mm,70mm){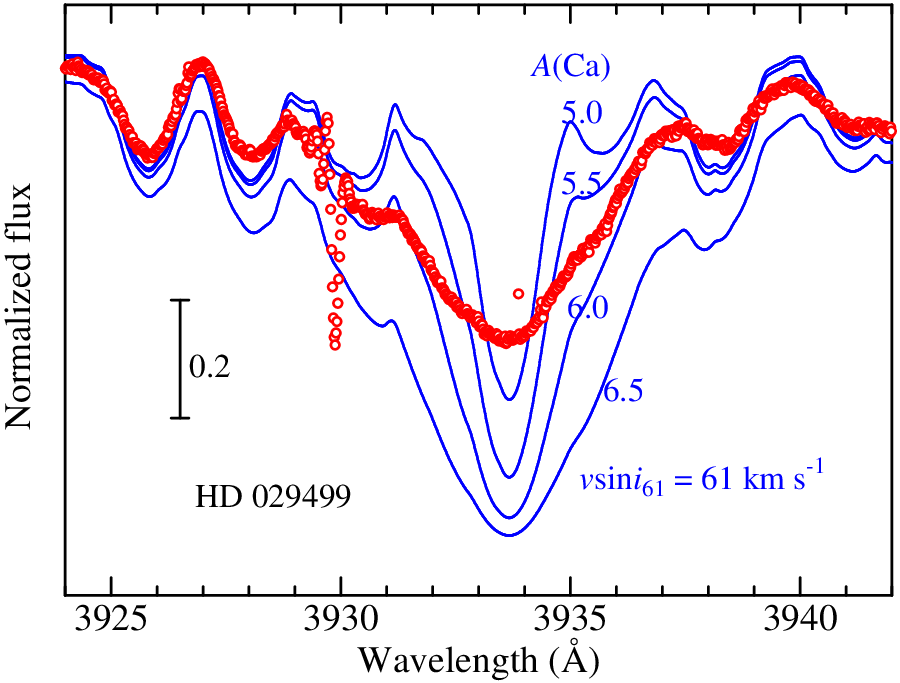}
  \end{center}
\caption{Demonstration of how the typical ``weak broad K line'' star
HD~29499 shows an anomalous Ca~{\sc ii}~3934 line profile.
Blue solid lines are theoretically synthesized profiles, which are computed 
for $A$(Ca) = 5.0--6.5 with a step of 0.5~dex and broadened with the standard 
$v\sin i_{61}$ of 61~km~s$^{-1}$ (actual projected rotational velocity 
for this star), while the observed profile is shown by red symbols (where  
the vertical offset and the global tilt are appropriately adjusted).
}
\end{figure}

\subsection{Ca abundances from Ca~{\sc i} 6162 and Ca~II 3934}

We are now ready to address the questions raised in section~1:
(i) How about the characteristics of the Ca abundances determined from 
Ca~{\sc ii} K line (especially in relation to the Am phenomenon)?
(ii) Are they consistent with those obtained from Ca~{\sc i} 6162 line?
In figure~8 are plotted the resulting Ca abundances from both lines ([Ca/H]$_{39}$, 
[Ca/H]$_{61}$) along with their differences against $T_{\rm eff}$ and $v\sin i_{61}$.

Let us discuss [Ca/H]$_{61}$ first. It can be seen from figure~8b that 
(i) the upper envelope of the distribution is [Ca/H]$_{61} \sim 0$ irrespective 
of $T_{\rm eff}$, (ii) Am stars are generally underabundant in Ca compared to 
normal stars, and (iii) the extent of deficiency (lower envelope of the distribution) 
tends to increase with a decrease in $T_{\rm eff}$. Furthermore, this deficiency 
of Ca has a systematic trend in terms of rotational velocity according to 
figure~8e, which shows that (iv) [Ca/H]$_{61}$ is nearly normal in rapid rotators
($v\sin i_{61} \gtsim 100$~km~s$^{-1}$) while (v) begins to progressively decrease
with a lowering of $v\sin i_{61}$ at $v\sin i_{61} \ltsim 100$~km~s$^{-1}$.     
These characteristics of [Ca/H]$_{61}$ seem to well fit the expectations from the 
diffusion theory, which predicts that photospheric Ca deficiency develops due to 
element segregation in the deeper stable radiative envelope of slow rotators 
(cf. the references quoted in section~1) and this anomaly tends to be enhanced 
towards lower $T_{\rm eff}$ (see, e.g., figure~14 in Richer et al. 2000).  
  
In contrast, the behaviors of [Ca/H]$_{39}$ are rather different (although some 
similar tendency to that of [Ca/H]$_{61}$ is also seen in the qualitative sense): 
[Ca/H]$_{39}$ tends to be lower than [Ca/H]$_{61}$ and the difference tends to grow
towards lower $T_{\rm eff}$ as displayed in figure~8c, which shows that 
[Ca/H]$_{39}$ is considerably below [Ca/H]$_{61}$ by $\sim 1$~dex or more
in late A stars (especially Am stars) at $T_{\rm eff} \ltsim 8000$~K while 
[Ca/H]$_{39}$ is nearly on the same order of (or slightly below) [Ca/H]$_{61}$ 
at early A stars ($T_{\rm eff} \ltsim 8000$~K). As a result, [Ca/H]$_{39}$
progressively declines with a decrease of $T_{\rm eff}$ for both Am and normal 
stars (cf. figure~8a). Meanwhile, this discrepancy seems to be independent of
$v\sin i_{61}$ (figure~8f). 

Also, with a suspicion that this $T_{\rm eff}$-dependent Ca~{\sc ii}--Ca~{\sc i} 
abundance difference might be due to a non-LTE effect, test non-LTE calculations 
were carried out for representative cases of different ($T_{\rm eff}$, $A$(Ca))
combinations. However, the resulting non-LTE corrections turned out to be too 
small to explain the problem. These supplementary calculations are separately 
described in appendix~1. 
 
Accordingly, the following speculations may as well be made based on what has 
been elucidated regarding the characteristics of [Ca/H]$_{61}$ and [Ca/H]$_{39}$.
\begin{itemize}
\item
Ca abundances derived from Ca~{\sc i} 6162 are considered to be reliable and 
thus [Ca/H]$_{61}$ represents the true photospheric abundance of calcium.
Therefore, the observed characteristics of [Ca/H]$_{61}$ (see (i)--(v) above)
may suggest that Ca is actually deficient in the atmosphere of Am stars, which 
stems from the atomic diffusion process operating in the stable radiative 
envelope of slow rotators.
\item
On the other hand, Ca abundances derived from Ca~{\sc ii} 3934 are probably 
not correct especially for late A-type stars ($T_{\rm eff} \ltsim 8000$~K) 
where [Ca/H]$_{39}$ seems to be considerably underestimated in comparison
with (presumably correct) [Ca/H]$_{61}$ by up to $\sim 1$~dex or more.
\item
However, the cause for this erroneous underestimation of [Ca/H]$_{39}$ is 
totally unknown. It may be possible that this has something to do with
the WBK phenomenon (cf. subsection~5.3), which is seen mainly in slowly rotating 
late A-type stars apparently showing conspicuously low [Ca/H]$_{39}$. 
Clarifying the reason why Ca~{\sc ii} K line shows such an unusual behavior 
in slow rotators of cool A stars ($T_{\rm eff} \ltsim 8000$~K) should be an 
important issue to be addressed.
\item
As to the question regarding the cause for the weakness of Ca~{\sc ii} K line 
in Am stars, which motivated this study, two compound factors seem to be 
responsible:  (1) Deficiency of Ca abundance in the photosphere possibly 
resulting from the diffusion process (as indicated by [Ca/H]$_{61}$) is 
naturally the one reason. (2) Yet, there is more than that. Some unknown
special mechanism of weakening Ca~{\sc ii} 3934 line so operates in late A stars
($T_{\rm eff} \ltsim 8000$~K) as to further decrease the strength of this line. 
This synergy effect would be the reason why classical Am stars
(mostly late A-type) show markedly weak Ca~{\sc ii} K line.     
\item
Accordingly, we would have to realize that some unknown aspect beyond our 
current comprehension (i.e., WBK phenomenon or extra weakening mechanism
of Ca~{\sc ii} K line) still exists regarding chemically peculiar stars 
(including Am stars) of upper main sequence. In this field of stellar 
physics, abundance anomaly caused by atomic diffusion has been the main stream 
interpretation for their spectrum peculiarity for almost a half century.   
Admittedly, this line of theoretical investigation is surely important 
and still to be pursued. In addition, however, it may be worth paying 
attention also to a possibility of non-classical atmospheric effect 
(e.g., stratification/ionization difference in slowly rotating stars 
compared to rapid rotators) such as previously considered by 
B\"{o}hm-Vitense (1976, 2006) for the cause of Am phenomenon.
\end{itemize}

\setcounter{figure}{7}
\begin{figure}
  \begin{center}
    \FigureFile(80mm,100mm){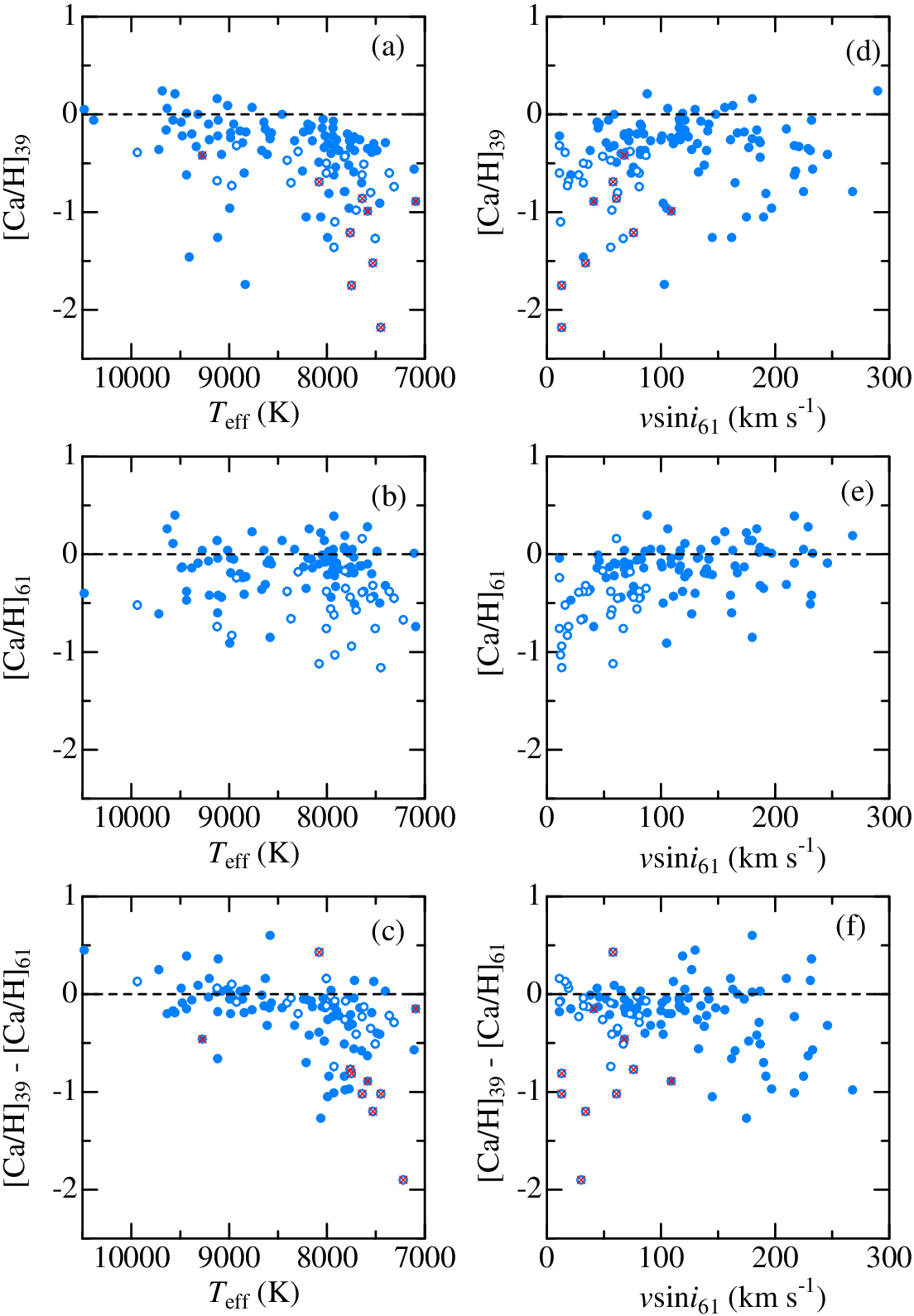}
  \end{center}
\caption{Comparison of two Ca abundances (relative to the reference star 
Procyon; cf. table~1) corresponding to Ca~{\sc ii} 3934 and Ca~{\sc i} 6162 lines
and their trends in terms of $T_{\rm eff}$ and $v\sin i_{61}$.
[Ca/H]$_{39}$, [Ca/H]$_{61}$, and [Ca/H]$_{39}-$[Ca/H]$_{61}$ are plotted
against $T_{\rm eff}$ in the left-hand panels (a--c), and against
$v\sin i_{61}$ in the right-hand panels (d--f). The meanings of the symbols
are the same as in figure~7.
}
\end{figure}

\subsection{Comparison with previous work}

Finally, it may be worth comparing our $W_{39}$ and $A_{39}$ results with 
the data derived by Guthrie (1987),who determined the equivalent widths of 
Ca~{\sc ii} K line and the resulting Ca abundances for a large number of 57
Am stars, though his work is based essentially on the literature data 
of $k$ and Sp(K) which were converted to $W_{39}$ by using the empirical relation.

Since 13 Am stars of $T_{\rm eff} \ltsim 8000$~K (\#65, \#75, \#77, \#86, \#94, 
\#95, \#106, \#107, \#109, \#111, \#112, \#118, and \#119) among his 57 sample 
stars are in common with our program stars, comparisons of Ca~{\sc ii} K-related 
results for these 13 objects are shown in figure~9. As can be seen from this figure,
a considerably large discrepancy is seen in the sense that Guthrie's (1987) 
$W_{39}$ as well as $A_{39}$ are systematically large (by a factor of $\ltsim 2$
for $W_{39}$ and by $\ltsim 1$~dex for $A_{39}$, even if excluding the data for 
WBK stars which are unreliable and probably underestimated; cf. footnote~3 ). 

The author has no idea about the reason for such a seriously large difference.
Since Guthrie's analysis is based on the published literature data of photometric 
$k$ index or K-line spectral type, its quantitative credibility may not be so high, 
especially for cool Am stars where contamination of neighboring 
metallic lines are significant.
On the other hand, North et al. (1997) stated that their Ca~{\sc ii} K line results 
for 23 Am stars based on spectrum synthesis (probably more reliable) are in good 
agreement with those of Guthrie (1987), though unfortunately North et al. did not 
publish any detailed data necessary for further check. If the conclusions of these 
previous investigations are correct, something might be wrong with the results 
of this study.

In any event, the problem is that research work on the Ca~{\sc ii} K line in 
A-type stars has been insufficient so far as mentioned in section~1. Consequently, 
further follow-up study on this subject is necessary, by which more light would 
be shed on this situation.

\setcounter{figure}{8}
\begin{figure}
  \begin{center}
    \FigureFile(50mm,100mm){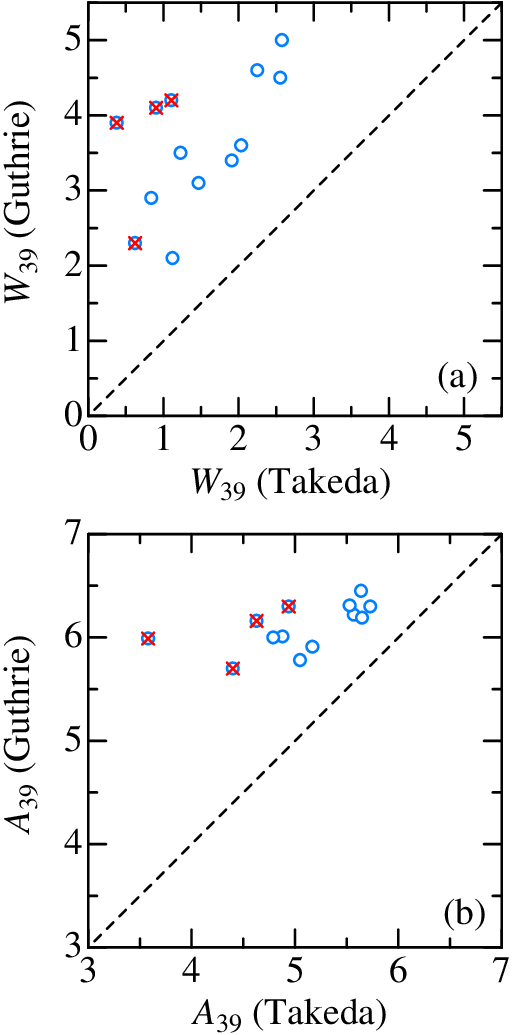}
  \end{center}
\caption{
Comparison of (a) $W_{39}$ (equivalent width of Ca~{\sc ii} 3934 in \AA)
and (b) $A_{39}$ (logarithmic Ca abundance) derived by Guthrie (1987)
based on Henry's (1969) narrow-band K-line photometry data
with those derived in this study, for 13 stars in common (all are
cool Am stars of $T_{\rm eff} \ltsim 8000$~K, among which 4 are
weak broad K line stars). The same meanings of symbols as in figure~3.
}
\end{figure}

\section{Summary and conclusion}

Despite that the strong Ca~{\sc ii} K line at 3934~\AA\ plays an important role 
in the spectral classification of upper main-sequence stars (its weakness 
is one of the criteria for Am stars), quantitative spectroscopic studies of 
this Ca~{\sc ii} 3934 line for a large number of A-type stars have been 
insufficient so far, and the cause of Ca~{\sc ii} K line weakening in relation 
to the Am phenomenon is not clarified yet. In order to address this issue,
studying this Ca~{\sc ii} K line along with the neutral Ca~{\sc i} lines 
for comparison should be worthwhile.  

About a decade ago, Takeda et al. (2009) carried out a synthetic 
spectrum-fitting analysis applied to the 6140--6170~\AA\ region for a large 
sample of 122 A-type main-sequence stars ($T_{\rm eff} \sim$~7000--10000~K) 
in a wide rage of rotational velocity ($v\sin i \sim$~10--300~km~s$^{-1}$). 
Since this region included Ca~{\sc i} lines, Ca abundances from Ca~{\sc i} 
lines ([Ca/H]$_{61}$) and Ca~{\sc i} 6162 line strengths ($W_{61}$) 
are already available for these sample stars.

In this investigation, the same spectrum synthesis technic was applied 
anew to the 3910--3955~\AA\ region (comprising Ca~{\sc ii} 3934)
to determine the Ca abundances from Ca~{\sc ii} 3934 ([Ca/H]$_{39}$)
and its line strengths ($W_{39}$) for these 122 program stars (including 
28 Am stars) by using the same observational data, in order to clarify 
(i) how the resulting Ca abundances are related with the Am phenomenon 
and (ii) whether [Ca/H]$_{39}$ and [Ca/H]$_{61}$ are consistent with each other.

It was confirmed that, for both Ca~{\sc i} and Ca~{\sc ii} lines, Ca line 
strengths in Am stars tend to be weaker and associated abundances are lower 
compared to non-Am stars at the same $T_{\rm eff}$, indicating a deficiency 
of Ca in the photosphere of Am stars which are mostly slow rotators. 
Since the upper envelope of [Ca/H]$_{61}$ vs. $T_{\rm eff}$ distribution 
is [Ca/H]$_{61} \sim 0$ (irrespective of $T_{\rm eff}$) for 
normal A stars, it may be reasonable to state that Ca~{\sc i} 6162 line 
yields correct Ca abundances.

Regarding the comparison between [Ca/H]$_{39}$ and [Ca/H]$_{61}$, 
while both are roughly consistent with each other for hotter stars 
($T_{\rm eff} \gtsim 8000$~K), the former tends to be lower (by up to $\sim 1$~dex or 
even more) than the latter for cooler A stars ($T_{\rm eff} \ltsim 8000$~K).
Accordingly, [Ca/H]$_{39}$ is likely to be erroneously underestimated in such
late A-type stars for some unknown reason.

In addition, an appreciable fraction of these cool A (mostly Am) stars
were found to show unusually anomalous Ca~{\sc ii} 3934 line feature 
(i.e., extraordinarily broad for its considerable weakness) which is hard to explain.
These ``Weak Broad K-line'' stars yielded apparently low Ca~{\sc ii} K line 
strengths and [Ca/H]$_{39}$ values; but they are subject to large uncertainties
(presumably underestimated) and thus unreliable, because theoretically 
synthesized spectrum could not fit the observed spectrum.

Accordingly, some special mechanism reducing the strength (and broaden the width) 
of Ca~{\sc ii} 3934 line seems to be involved at $T_{\rm eff} \ltsim 8000$~K 
where [Ca/H]$_{39}$ would be no more reliable. 

Atomic diffusion process (causing the deficit of Ca in the photosphere as a result of
element segregation in the deeper radiative envelope) is currently the promising 
explanation for the Am phenomenon, which seems to be consistent with the qualitative 
trend of [Ca/H]$_{61}$ in A-type stars. When it come to Ca~{\sc ii} 3934 K line, 
however, its considerable weakness in classical Am stars should not necessarily 
be attributed to only this element diffusion scenario; i.e., some independent 
mechanism weakening this resonance line (specific to slowly rotating Am stars) 
may also be working.

As a last remark, the equivalent widths of the Ca~{\sc ii} 3934 line measured 
in this study for cool Am stars (and the associated Ca abundances) appear to be 
in conflict (i.e., tend to be systematically smaller) with those of previous 
investigations. Further follow-up verification studies are to be desirably awaited. 

\bigskip

This research has made use of the SIMBAD database, operated by
CDS, Strasbourg, France. 
This investigation has also made use of the VALD database, operated at Uppsala 
University, the Institute of Astronomy RAS in Moscow, and the University of Vienna.
Part of the observational data of this work is taken from the stellar spectra 
archive of the UVES Paranal Observatory Project (ESO DDT Program ID 266.D-5655).

\appendix

\section{Non-LTE effect on Ca abundance determinations}

Although a number of papers have been published on the non-LTE line formation 
of Ca~{\sc ii} resonance lines at 3934~\AA\ (K) and 3968~\AA\ (H) for solar-type 
(F-, G-, and K-type) stars because core emission feature of these doublet lines 
is a good indicator of chromospheric activity, non-LTE effect on Ca abundance
determinations in A-type stars (especially in relation to the Ca~{\sc ii} K line) 
has been scarcely studied so far. As far as the author knows, it is probably 
only the following two investigations that are available at present:
(i) Snijders (1975) carried out non-LTE calculations for Ca~{\sc ii} in the 
$T_{\rm eff}$ range of 10000--15000~K (hot A and late B stars). His results are
described in Leckrone (1976), who stated that the non-LTE effect depends on 
the treatment of UV flux. (ii) St\"{u}renburg (1993) determined the non-LTE
corrections for Ca~{\sc i} and Ca~{\sc ii} lines for A-type stars of
$T_{\rm eff} \sim$~7500--9000~K in his abundance analysis of 
$\lambda$~Bootis stars. His results indicated that the non-LTE correction is
line-dependent but quantitatively insignificant because its signs are
negative and positive around zero.
 
On this occasion, the author conducted test non-LTE calculations for Ca~{\sc i} 6162 
and Ca~{\sc ii} 3934 lines in scope of A-type main-sequence stars, in order to 
examine whether or not the results of this study are significantly affected 
by the non-LTE effect.
The atomic model and the computational procedure are the same as adopted
by Takeda et al. (2010; cf. appendix~B therein) in their non-LTE simulation
of solar Ca~{\sc ii} K line core emission profiles.
The statistical-equilibrium calculations were carried out for 
three models with $T_{\rm eff}$ = 7000, 8500, and 10000~K
(each with the same solar-metallicity and $\log g=4$) and 
$v_{\rm t} = 2$~km~s$^{-1}$.
 
In figure 10 are shown the $l_{0}^{\rm NLTE}(\tau)/l_{0}^{\rm LTE}(\tau)$ 
(the NLTE-to-LTE line-center opacity ratio; nearly equal to 
$\simeq b_{\rm l}$) and $S_{\rm L}(\tau)/B(\tau)$ (the ratio of 
the line source function to the Planck function; nearly equal to 
$\simeq b_{\rm u}/b_{\rm l}$), where $b_{\rm l}$ and $b_{\rm u}$ are 
the non-LTE departure coefficients for the lower and upper levels, 
respectively) for each of the the Ca~{\sc i} 4$^{3}$P$^{\rm o}$--5$^{3}$S 
transition (relevant to Ca~{\sc i} 6162 of multiplet 3) and
Ca~{\sc ii} 4$^{2}$S--4$^{2}$P$^{\rm o}$ transition (corresponding to
Ca~{\sc ii} 3934 of multiplet 1).
Also, the LTE and non-LTE equivalent widths for the 
Ca~{\sc i} 6162 and Ca~{\sc ii} 3934 lines along with the corresponding 
non-LTE abundance corrections were computed for each case, which are 
summarized in table 3.

An inspection of figure 10 and table 3 reveals the following characteristics
regarding the non-LTE effect for these two Ca lines.
\begin{itemize}
\item
Generally speaking, non-LTE departure is restricted to high
atmosphere, and the non-LTE corrections ($\Delta$) are insignificant
for both lines ($\ltsim 0.1$~dex in most cases).
\item
Regarding Ca~{\sc ii} 3934, $\Delta_{39}$ values are negative (corresponding to 
non-LTE line strengthening reflecting $S_{\rm L}/B < 1$)
but their extents are almost negligible, except for the saturated line case 
of high forming depth ($W_{39} \ltsim 200$~m\AA\ for the t100g40 model with 
$A = 4.36$ in table~3) corresponding to the flat part of the curve of growth.
\item
In contrast, behaviors of $\Delta_{61}$ for Ca~{\sc i} 6162 are more complex,
in the sense that its sign can be positive as well as negative.
This is due to the fact that both opposite effects of (1) underpopulation of the 
lower level (line weakening) and (2) dilution of $S_{\rm L}$ (line 
strengthening) act in the formation of this line. Again, $\Delta_{61}$
becomes non-negligible for the cases of saturated lines 
($W_{61} \sim$~100--200~m\AA; $A = 6.36$ case for t070g40 and t085g40 models).
\item
In any event, non-LTE effect is not quantitatively significant 
as far as the Ca lines used in this investigation for the analysis of
A-type stars are concerned, which means that the assumption of LTE is
reasonable in the practical sense. Also, the resulting trend of
$\Delta$ values in table~3 indicates that these (conventionally evaluated) 
non-LTE corrections are unable to explain the serious discrepancy
between [Ca/H]$_{39}$ and [Ca/H]$_{61}$ mentioned in subsection~5.4.
\end{itemize}

\setcounter{figure}{9}
\begin{figure}
  \begin{center}
    \FigureFile(80mm,100mm){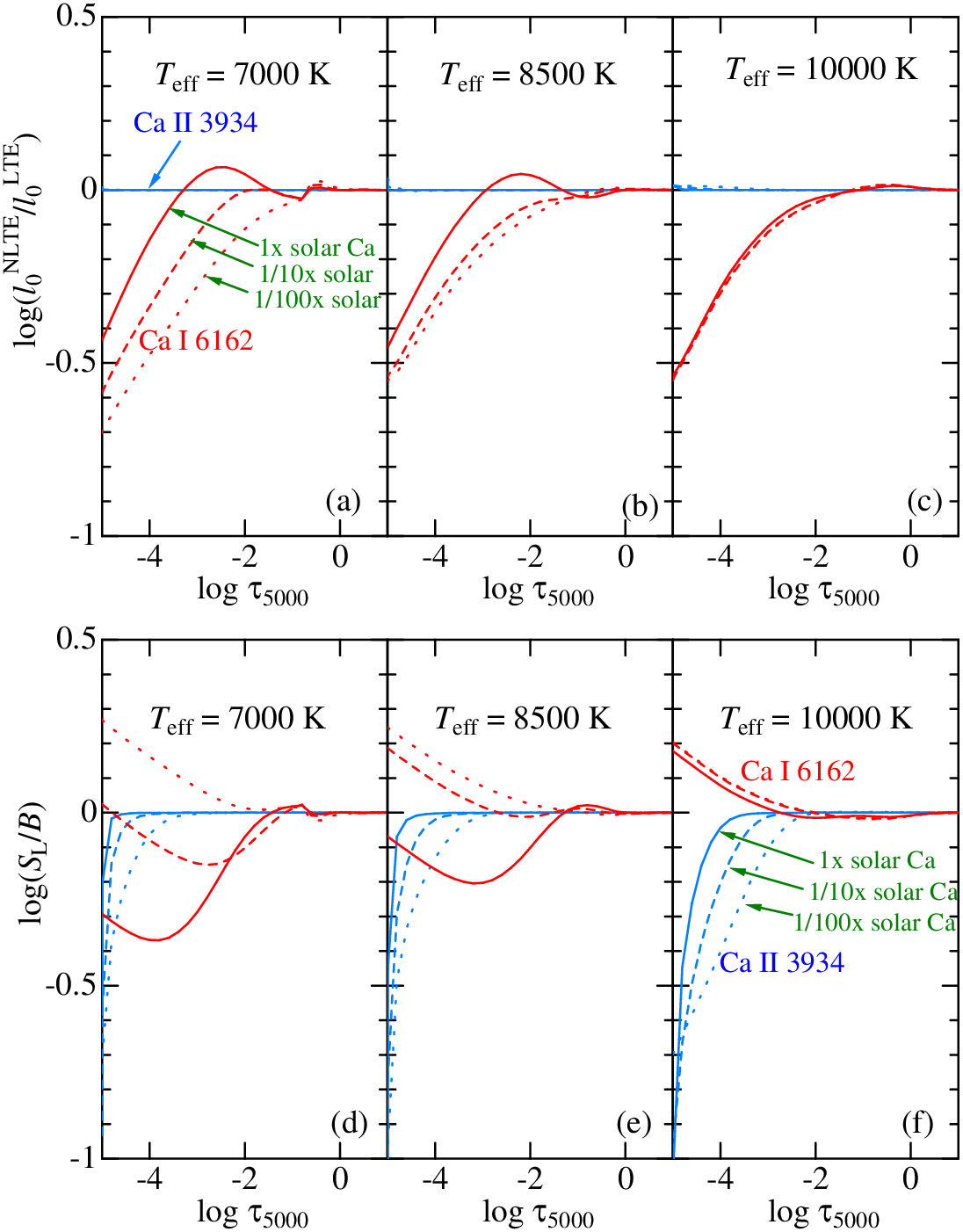}
  \end{center}
\caption{
The NLTE-to-LTE line-center opacity ratio (upper panels a--c) and 
the ratio of the line source function ($S_{\rm L}$) 
to the local Planck function ($B$) (lower panels d--f)  
for the Ca~{\sc i} 4$^{3}$P$^{\rm o}$--5$^{3}$S transition
of Ca~{\sc i}~6162 (red lines)
and the Ca~{\sc ii} 4$^{2}$S--4$^{2}$P$^{\rm o}$ transition of
Ca~{\sc ii} 3934 (blue lines), plotted against the continuum
optical depth at 5000~\AA. 
Computations were done on three $\log g = 4.0$ solar-metallicity 
models with different $T_{\rm eff}$ of 7000~K (left panels a, d), 
8500~K (middle panels b, e), and 10000~K (right panels c, f),
for three Ca abundances of $A$(Ca) = 6.36 (solar), 5.36 (1/10$\times$ solar)
and 4.36 (1/100$\times$ solar) as depicted in solid, dashed, and dotted lines,
respectively.
}
\end{figure}

\section{Effect of abundance stratification on the Ca~II K line profile}

As described in subsection~5.3, the existence of `Weak Broad K line'' (WBK) 
stars, which are mainly seen in slowly rotating cool Am stars, was an
embarrassing finding. How could such an extraordinary profile be possible?

When it comes to anomalous Ca~{\sc ii} 3934 line profile in chemically peculiar 
stars, what comes to our mind is the case of some cool Ap stars
showing a peculiar K line profile with sharp core and broad wing.
This unusual Ca~{\sc ii} K line shape is nowadays considered to be due to 
the stratified Ca abundance in the atmosphere (i.e., Ca atoms settle out 
towards deeper subphotospheric layer, leading to a deficiency in the 
upper atmosphere and an overabundance in the deep photosphere).
Actually, an appropriately adjusted stratified model of Ca distribution 
can reproduce the observed K line profile of such Ap stars remarkably well 
(see, e.g., figure~3 in Ryabchikova 2014). 

This is demonstrated in figure~11, where two Ca~{\sc ii} K lines profiles
computed for two cases of Ca distribution (homogeneous case and stratified case)
are compared with each other. Here, the atmospheric parameters and the Ca 
abundance profile were chosen by consulting figure~4 of Ryabchikova (2014). 
It can be seen from this figure that the stratification of Ca atoms can produce
such a characteristic profile (with wider wing and sharper core in comparison 
with normal profile) as actually observed in Ap stars (e.g., 10~Aql).

However, these examples in figure~11 correspond to the case of damping-dominated 
strong line and do not represent the weak Ca~{\sc ii} K line
as observed in WBK stars ($W_{39} \ltsim 1000$~m\AA; cf. figure~3a).
Therefore, simulations were carried out to examine whether it is possible 
to produce weak but broad profile by adequately adjusting the stratified 
Ca abundance distribution. The results are depicted in figure~12 (see the 
caption of this figure for the details of this test calculation). 
Although we refrain from going into in-depth discussion, one important 
conclusion elucidated from figure~12 should be mentioned: In case where 
the line strength is considerably weak ($\log W \ltsim 3$) compared to those of 
normal A stars, the line profile can not be made significantly different
from that of the ordinary homogeneous case (cf. the left part of each panels b--d 
and f--h in figure~12, where we can see that symbols are asymptotically
merged with the red line), no matter how the distribution of Ca stratification 
is changed (as far as the present modeling is concerned).
Accordingly, it may be concluded that the depth-dependent Ca abundance
distribution can not be an explanation for the WBK phenomenon.  

\setcounter{figure}{10}
\begin{figure}
  \begin{center}
    \FigureFile(60mm,70mm){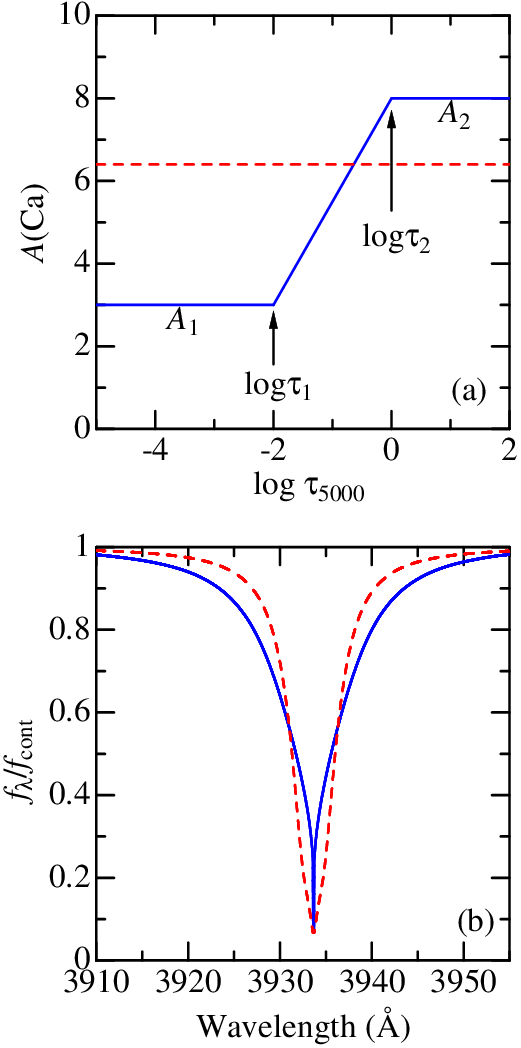}
  \end{center}
\caption{
Demonstrative Ca~{\sc ii} 3934 profile simulation carried out on  
the representative $T_{\rm eff} = 7500$~K and $\log g = 4.0$ model for 
two different Ca distributions as illustrated in panel (a): 
(i) depth-independent homogeneous Ca abundance of $A = 6.4$ (red dashed line) 
and (ii) stratified Ca abundances with $A_{1} = 3.0$, $A_{2} = 8.0$,
$\log \tau_{1} = -2$, and $\log \tau_{2} = 0$ (blue solid line).
The resulting profiles for the two cases are depicted in panel (b).
}
\end{figure}

\setcounter{figure}{11}
\begin{figure}
  \begin{center}
    \FigureFile(80mm,160mm){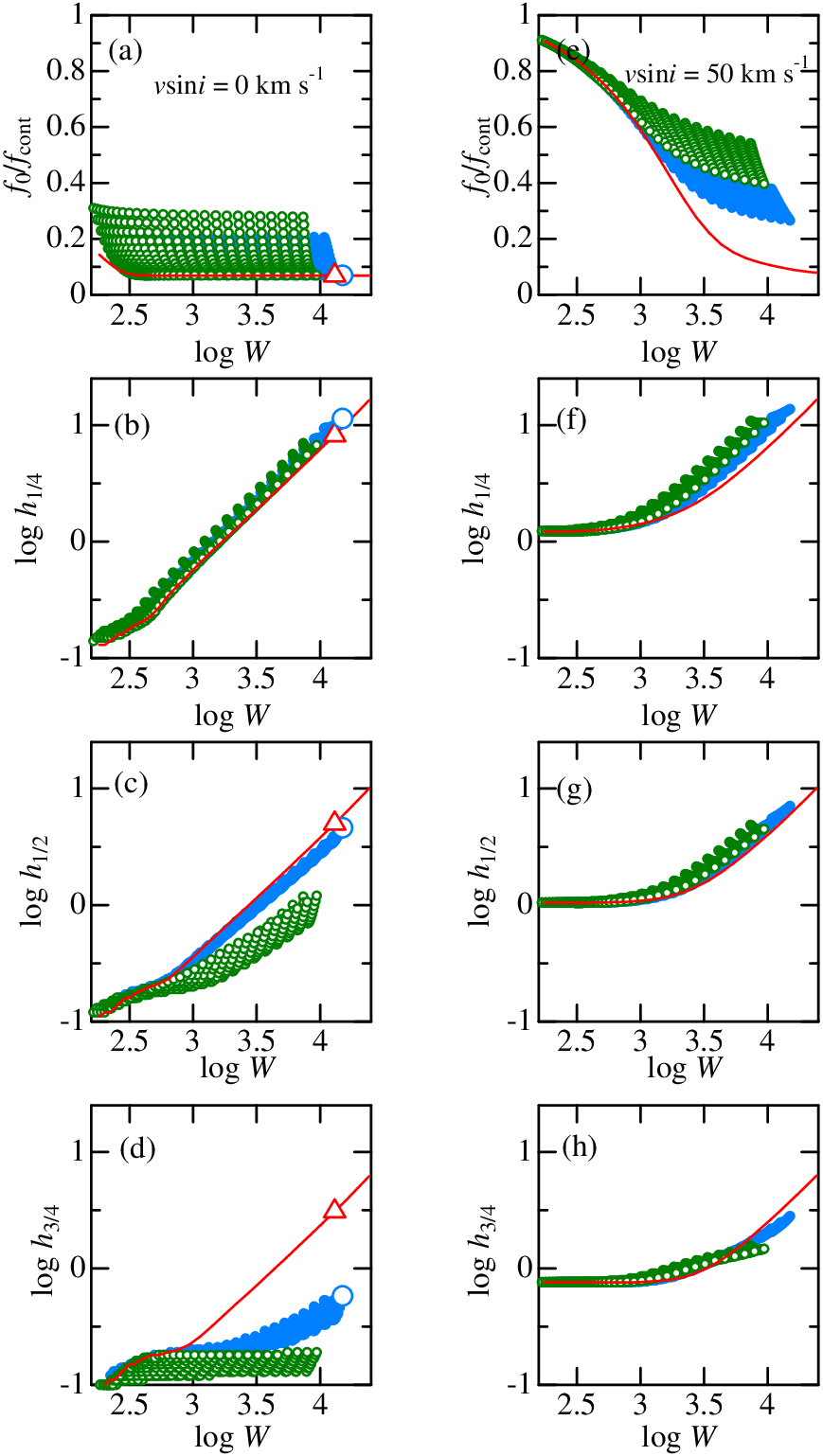}
  \end{center}
\caption{
Characteristics of theoretical Ca~{\sc ii} 3934 line profiles simulated for 
many stratified Ca abundance distributions where the combinations of 
$(A_{1}, A_{2})$ are variously changed in the ranges of $A_{1}=$~0--3 
and $A_{2}=$~3--8 (with a step of 0.2~dex), while two cases are assumed
for the critical optical depths: (i) gradual gradient with 
$(\log \tau_{1}, \log \tau_{2}) = (-2, 0)$ and
(ii) steeper gradient with $(\log \tau_{1}, \log \tau_{2}) = (-1, 0)$. 
In panels (a/e), (b/f), (c/g), and (d/h) are plotted the line-center 
residual flux ($f_{0}/f_{\rm cont}$), 1/4-maximum width ($h_{1/4}$), 
half-maximum width ($h_{1/2}$),and 3/4-maximum width ($h_{1/3}$)  
against the correspondent equivalent width ($W$), where $h$ is in unit of
\AA\ and $W$ is in m\AA.
The left-hand panels (a--d) and right-hand panels (e--h) correspond to
the unbroadened profiles ($v\sin i = 0$~km~s$^{-1}$) and rotationally 
broadened profiles with $v\sin i = 50$~km~s$^{-1}$, respectively.
Case (i) and case (ii) results are expressed in blue filled circles  
and green open circles, respectively. The results for the case of 
homogeneous Ca abundance ($A_{1}=A_{2}$) are shown by red solid lines
for comparison. Two large open symbols in the left-hand panels (a--d) 
correspond to two example cases shown in figure~11.  
}
\end{figure}

\onecolumn

\setcounter{table}{2}
\begin{table}[h]
\begin{minipage}{180mm}
\caption{NLTE abundance corrections for Ca~{\sc ii} 3934 and Ca~{\sc i} 6162 lines.}
\scriptsize
\begin{center}
\begin{tabular}
{cccccccc}
\hline \hline
Code & $v_{\rm t}$ & $A^{\rm a}$ & ($W^{\rm LTE}$) & $W^{\rm NLTE}$ & $A^{\rm N}$ & $A^{\rm L}$ & $\Delta$ \\
     &  (km~s$^{-1}$) & (dex)   & (m\AA) & (m\AA) & (dex)   & (dex) & (dex) \\
\hline
\multicolumn{8}{c}{[Ca~{\sc ii} 3933.66 line]} \\
t070g40 & 2.0 & 6.360 & (8550.67)& 8609.94 & 6.360 & 6.365 &$-0.005$ \\
t070g40 & 2.0 & 5.360 & (2703.96)& 2722.70 & 5.361 & 5.366 &$-0.005$ \\
t070g40 & 2.0 & 4.360 & ( 866.96)&  872.97 & 4.360 & 4.367 &$-0.007$ \\
t085g40 & 2.0 & 6.360 & (3126.08)& 3147.75 & 6.360 & 6.366 &$-0.006$ \\
t085g40 & 2.0 & 5.360 & ( 995.41)& 1002.31 & 5.359 & 5.366 &$-0.007$ \\
t085g40 & 2.0 & 4.360 & ( 345.14)&  356.45 & 4.359 & 4.395 &$-0.036$ \\
t100g40 & 2.0 & 6.360 & ( 805.38)&  812.83 & 6.359 & 6.369 &$-0.010$ \\
t100g40 & 2.0 & 5.360 & ( 297.85)&  310.46 & 5.360 & 5.405 &$-0.045$ \\
t100g40 & 2.0 & 4.360 & ( 167.49)&  178.65 & 4.362 & 4.518 &$-0.156$ \\
\hline
\multicolumn{8}{c}{[Ca~{\sc i} 6162.17 line]} \\
t070g40 & 2.0 & 6.360 & ( 141.91)&  164.06 & 6.359 & 6.652 &$-0.293$ \\
t070g40 & 2.0 & 5.360 & (  79.07)&   84.72 & 5.361 & 5.451 &$-0.090$ \\
t070g40 & 2.0 & 4.360 & (  20.70)&   18.71 & 4.360 & 4.306 &$+0.054$ \\
t085g40 & 2.0 & 6.360 & (  81.66)&   92.47 & 6.360 & 6.540 &$-0.180$ \\
t085g40 & 2.0 & 5.360 & (  22.44)&   20.75 & 5.359 & 5.317 &$+0.042$ \\
t085g40 & 2.0 & 4.360 & (   2.74)&    2.43 & 4.355 & 4.302 &$+0.053$ \\
t100g40 & 2.0 & 6.360 & (   7.60)&    7.71 & 6.359 & 6.366 &$-0.007$ \\
t100g40 & 2.0 & 5.360 & (   0.82)&    0.82 & 5.347 & 5.351 &$-0.004$ \\
t100g40 & 2.0 & 4.360 & (   0.08)&    0.08 & 4.441 & 4.446 &$-0.005$ \\
\hline
\end{tabular}
\end{center}
Note. 
Code ``t$aaa$g$bb$'' denotes the model (solar metallicity model with 
[Fe/H] = 0.0) with $T_{\rm eff} = aa \times 100$ and $\log g = bb / 10$.
Calculations for each model were made with the microturbulence 
$v_{\rm t} = 2$~km~s$^{-1}$ for three assigned Ca abundances (logarithmic 
number abundances in the usual normalization of $A_{\rm H} = 12.00$): 
$A_{i}^{\rm a}$ ($\equiv$ 6.36  + [Ca/H]$_{i}$)
($i$ = 1, 2, 3), where [Ca/H]$_{1}$, [Ca/H]$_{2}$, and [Ca/H]$_{3}$ 
are 0.0, $-1.0$, and $-2.0$, respectively.
$W^{\rm LTE}$ and $W^{\rm NLTE}$ are the resulting theoretical 
LTE and NLTE equivalent widths (in m\AA) corresponding to 
the assigned $A^{\rm a}$, respectively.
Based on such calculated non-LTE equivalent width ($W^{\rm NLTE}$),
two kinds of Ca abundances were inversely computed for the cases of
NLTE ($A^{\rm N}$) and LTE ($A^{\rm L}$), from which the non-LTE
abundance correction was eventually evaluated as the difference of
these two, $\Delta$ ($\equiv A^{\rm N} - A^{\rm L}$).
\end{minipage}
\end{table}

\newpage

\begin{center}
{\Large {\bf Erratum: Behaviors of Ca II K line in A-type stars}}\\
\vspace{2mm}
{\large {\it (2024 October 14, by Yoichi Takeda)}}
\end{center}

\vspace{5mm}
\small

In the article [Stars and Galaxies, Vol.~3, id.~1 (2020)], Ca abundances 
of 122 A-type stars were determined from the Ca~{\sc ii} K line at 3933.68~\AA\
and Ca~{\sc i} line at 6162.17~\AA.
Although the analysis was done under the assumption of LTE (Local Thermodynamical
Equilibrium), a supplementary discussion on the non-LTE effect for these two Ca lines 
was briefly presented in appendix~1 therein, based on statistical equilibrium 
calculations carried out for selected models of representative 
parameters.  

It has recently revealed, however, that part of the results (non-LTE corrections 
$\Delta$ and non-LTE abundances $A^{\rm N}$) derived therein was not correct  
because of an inadvertently erroneous treatment in the non-LTE simulation code. 
Specifically, the overionization effect of Ca~{\sc ii} atoms (acting to weaken 
Ca~{\sc ii} lines or shifting $\Delta$ towards the positive direction) was 
underestimated by this mistake, and this error tends to become more 
significant with an increase in $T_{\rm eff}$ (as the dominant ionization 
stage of Ca atoms changes from Ca~{\sc ii} to Ca~{\sc iii}).

Actually, this non-LTE effect (Ca~{\sc ii} overionization) can be quite important 
for early A-type stars of higher $T_{\rm eff}$, which would also leads to an 
underpopulation of Ca~{\sc i} in line with Ca~{\sc ii} parent terms.  

Therefore, the whole analysis of appendix~1 was redone by conducting new 
corrected non-LTE calculations. The updated results are shown figure~13 (behaviors 
of non-LTE line opacity and line source function) and table~4 (non-LTE abundances
and corrections), which should be compared with figure~10 and table~3 of the 
original article. The following characteristics are seen from this comparison.
\begin{itemize}
 \item
For both of the Ca~{\sc ii} 3934 and Ca~{\sc i} 6162 lines, while the 
results remain the same unchanged for the $T_{\rm eff} = 7000$~K case, some 
noticeable changes begin to emerge at $T_{\rm eff} = 8500$~K,  and 
considerable differences are manifestly seen at $T_{\rm eff} = 10000$~K.  
\item
Accordingly, in contrast to the old (incorrect) results where LTE 
practically holded for these Ca lines irrespective of $T_{\rm eff}$,
the non-LTE line-weakening effect due to overionization  
becomes significant at the high-$T_{\rm eff}$ regime.
\item
This marked difference is naturally reflected in the non-LTE corrections 
($\Delta_{3934}$ and $\Delta_{6162}$) for the $T_{\rm eff} = 10000$~K case, 
for which the previously obtained insignificantly small values 
cf. table~3) are changed to considerably large positive corrections 
($\sim +0.4$~dex; cf. table~4) by new calculations.
\end{itemize}

Consequently, the previous conclusion of appendix~1 (the non-LTE corrections
are $|\Delta| \ltsim$~0.1--0.2~dex at most and thus insignificant for both 
Ca~{\sc i} 6162 and Ca~{\sc ii} 3934 lines) was not necessarily correct,
which should be restated as ``$\Delta_{3934}$ as well as $\Delta_{6162}$ 
can be significantly large to be $\sim +0.4$~dex for early A-type stars 
with $T_{\rm eff}$ as high as $\sim 10000$~K (though $\Delta$s still remain 
insignificant in the lower $T_{\rm eff}$ regime)''.
It should be noted, however, that this revision of non-LTE corrections 
does not help to solve the problem of $A_{39}$ vs. $A_{61}$ discrepancy 
observed in late A-type stars.

\onecolumn

\setcounter{figure}{12}
\begin{figure}
\begin{minipage}{180mm}
  \begin{center}
    \FigureFile(90mm,90mm){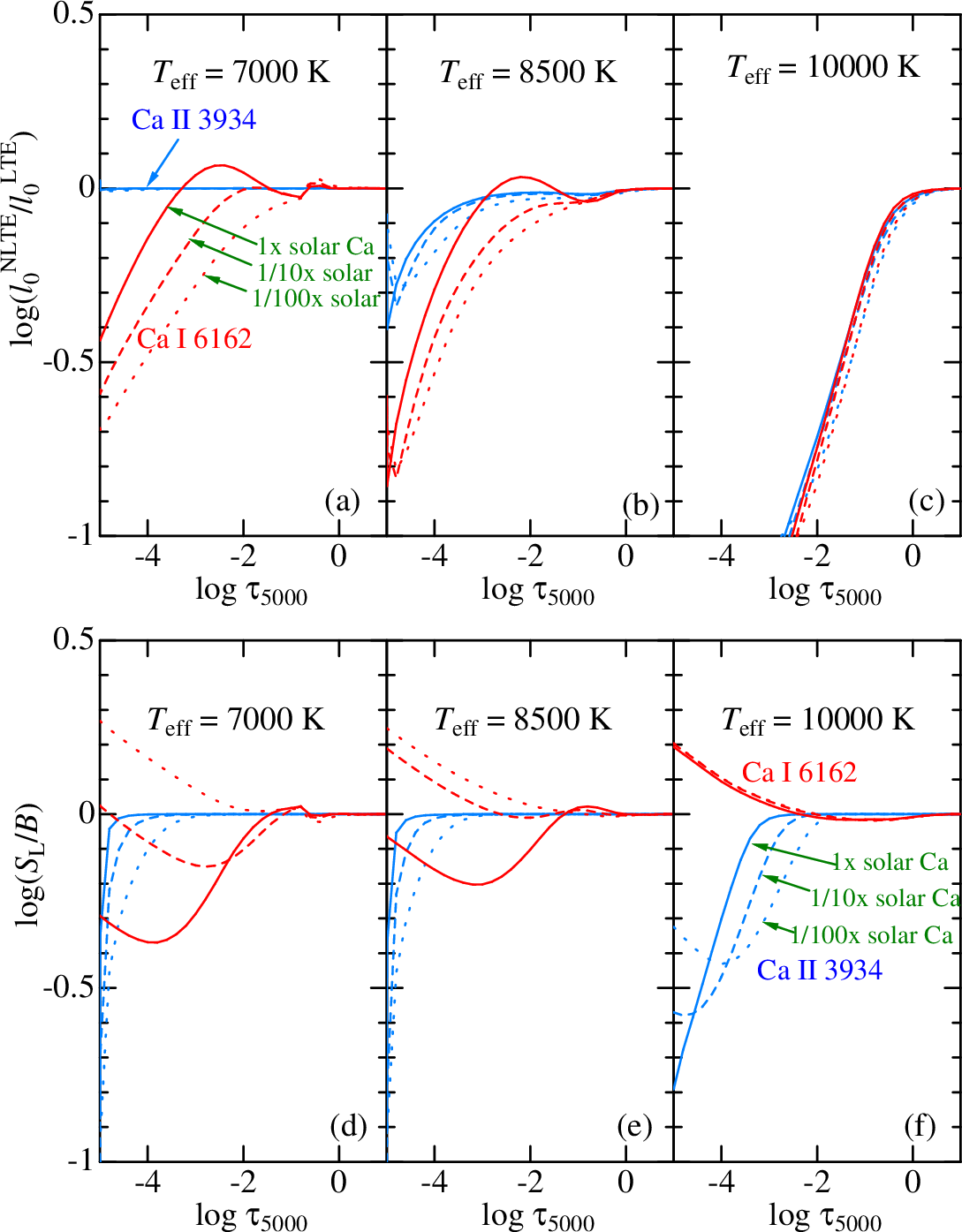}
  \end{center}
\caption{
Behaviors of revised non-LTE line opacity and line source function based on 
new corrected non-LTE calculations. This figure should be compared with figure~10 
of the original article. See the caption therein for more details. 
}
\end{minipage}
\end{figure}

\setcounter{table}{3}
\begin{table}[h]
\begin{minipage}{180mm}
\caption{Revised NLTE abundance corrections for Ca~{\sc ii} 3934 and Ca~{\sc i} 6162 lines.}
\scriptsize
\begin{center}
\begin{tabular}
{cccccccc}
\hline \hline
Code & $v_{\rm t}$ & $A^{\rm a}$ & ($W^{\rm LTE}$) & $W^{\rm NLTE}$ & $A^{\rm N}$ & $A^{\rm L}$ & $\Delta$ \\
     &  (km~s$^{-1}$) & (dex)   & (m\AA) & (m\AA) & (dex)   & (dex) & (dex) \\
\hline
\multicolumn{8}{c}{[Ca~{\sc ii} 3933.66 line]} \\
t070g40 & 2.0 & 6.360 & (8550.67)& 8609.94 & 6.360 & 6.365 &$-0.005$ \\
t070g40 & 2.0 & 5.360 & (2703.96)& 2722.70 & 5.360 & 5.366 &$-0.006$ \\
t070g40 & 2.0 & 4.360 & ( 866.96)&  872.97 & 4.359 & 4.367 &$-0.008$ \\
t085g40 & 2.0 & 6.360 & (3126.08)& 3076.10 & 6.359 & 6.346 &$+0.013$ \\
t085g40 & 2.0 & 5.360 & ( 995.41)&  977.24 & 5.360 & 5.343 &$+0.017$ \\
t085g40 & 2.0 & 4.360 & ( 345.14)&  345.94 & 4.359 & 4.361 &$-0.002$ \\
t100g40 & 2.0 & 6.360 & ( 805.38)&  539.51 & 6.359 & 5.989 &$+0.370$ \\
t100g40 & 2.0 & 5.360 & ( 297.85)&  225.94 & 5.361 & 4.985 &$+0.376$ \\
t100g40 & 2.0 & 4.360 & ( 167.49)&  143.55 & 4.360 & 3.930 &$+0.430$ \\
\hline
\multicolumn{8}{c}{[Ca~{\sc i} 6162.17 line]} \\
t070g40 & 2.0 & 6.360 & ( 141.91)&  164.06 & 6.359 & 6.651 &$-0.292$ \\
t070g40 & 2.0 & 5.360 & (  79.07)&   84.72 & 5.361 & 5.451 &$-0.090$ \\
t070g40 & 2.0 & 4.360 & (  20.70)&   18.71 & 4.361 & 4.306 &$+0.055$ \\
t085g40 & 2.0 & 6.360 & (  81.66)&   91.20 & 6.360 & 6.518 &$-0.158$ \\
t085g40 & 2.0 & 5.360 & (  22.44)&   20.04 & 5.360 & 5.300 &$+0.060$ \\
t085g40 & 2.0 & 4.360 & (   2.74)&    2.27 & 4.360 & 4.277 &$+0.083$ \\
t100g40 & 2.0 & 6.360 & (   7.60)&    3.16 & 6.360 & 5.958 &$+0.402$ \\
t100g40 & 2.0 & 5.360 & (   0.82)&    0.30 & 5.356 & 4.924 &$+0.432$ \\
t100g40 & 2.0 & 4.360 & (   0.08)&    0.03 & 4.422 & 3.923 &$+0.499$ \\
\hline
\end{tabular}
\end{center}
This table should be compared with table~3 of the original article. 
See the note therein for more details. 
\end{minipage}
\end{table}

\end{document}